# Human-Centered Human-AI Interaction (HC-HAII):
# A Human-Centered AI Perspective

Wei Xu

HCAI Labs, California, USA

**Abstract**: This chapter systematically promotes an emerging interdisciplinary field of human-artificial intelligence interaction (human-AI interaction, HAII) from a human-centered AI (HCAI) perspective. It introduces a framework of human-centered HAII (HC-HAII). HC-HAII places humans at the core of HAII research and applications, emphasizing the importance of adopting a human-centered approach over a technology-centered one. The chapter presents the HC-HAII methodology, including human-centered methods, process, interdisciplinary teams, and multi-level design paradigms. It also highlights key research challenges and future directions. As the first chapter, this chapter also provides a structural overview of this book, which brings together contributions from an interdisciplinary community of researchers and practitioners to advance the theory, methodology, and applications of HCAI in diverse domains of HAII. The purpose of this chapter is to provide a fundamental framework for this book, centered on HAII research and applications based on the HCAI approach, which will pave the way for the content of subsequent chapters.

## 1 Introduction

In the personal computer (PC) era, people's daily work and life mainly interacted with non-artificial intelligence (AI)-based computing systems (including products, applications, services) (Xu, 2003). In the AI era, AI systems (including products, applications, and services) have gradually entered people's daily work and life, allowing them to interact with these systems (i.e., human-AI interaction, HAII). However, the new features of AI technology make the interaction and user experience different from the traditional human-computer interaction that is primarily based on non-AI computing systems. This transition brings a series of new challenges, urging us to adopt new thinking and new approaches to research, design, and optimize this new type of interaction between humans and AI systems.

This chapter first discusses a series of new features of human-AI interaction and the new type of human-machine relationship brought by AI technology. Based on a human-centered AI (HCAI) approach, this chapter proposes a conceptual framework of human-centered HAII (HC-HAII). It elaborates on the aims, scope, key issues, methods, process, and design paradigms of the HC-HAII. The chapter also discusses the significance of conducting HAII practice from the HCAI perspective. The purpose of promoting HC-HAII is to ensure that the design, development, deployment, and use of AI systems enhance and empower human capabilities and values when humans interact with AI systems. HC-HAII extends beyond technical efficiency to embrace human needs, ethical alignment, and sociotechnical integration, delivering human-centered AI systems. Finally, this chapter highlights the structure of this book.

## 2 Emergence of human-AI interaction
### 2.1 Comparison of human interaction with AI systems and non-AI systems

Based on machine learning, big data, and other technologies, AI systems can be developed with autonomous features. Depending on the degree of autonomy, an AI system (including single or multiple AI agents) can have some cognitive abilities like humans (perception, learning, reasoning, etc.), autonomously performing certain tasks in specific scenarios. It may exhibit certain adaptability to unpredictable environments and can autonomously complete tasks in some unanticipated scenarios that previous automation technology could not complete (den Broek et al., 2017; Kaber, 2018; Rahwan et al., 2019; Xu, 2020, 2021).

Human-computer interaction (HCI) is a cross-disciplinary field that emerged in the PC era. It primarily studies the interaction between humans and non-AI computing systems. Table 1 compares basic characteristics of human interaction with non-AI systems and AI systems (Xu & Ge, 2020; Xu, 2020, 2021). Although current AI technology cannot yet fully realize all these new features, AI systems will gradually acquire these capabilities with continued advances in AI.

Table 1   Comparison between human-non-AI interaction vs. human-AI interaction

| Characteristics | Human-non-AI interaction (the PC era) | Human-AI interaction (the AI era) |
|---|---|---|
| Examples | Washing machines, elevators, and automated production lines | Intelligent decision-making systems, autonomous driving cars, and intelligent robots |
| Machine intelligence and behavior | Do not possess machine intelligence; behave according to predetermined fixed algorithms, logic, and rules | Varied levels of machine intelligence; varied levels of human-like cognitive abilities (e.g., learning, self-execution); unique machine behavior, learning, and evolution; machine behavior is unpredictable |
| Machine's role | A tool to assist human work | A potential collaborator for collaborative interaction with humans (based on the level of machine intelligence). |
| Machine output | Deterministic | Indeterministic and potentially biased |
| Human-machine relationship | Human-machine interaction | Human-AI collaborative interaction |
| Human-like sensing ability | Limited | Yes (with advanced technologies) |
| Human-like cognitive abilities (pattern recognition, learning, reasoning, etc.) | No | Yes (Abilities vary across design scenarios) |
| Human-like self-executing ability | No (Require human manual activation and intervention according to predefined rules | Yes (Perform operations independently in specific situations; abilities vary across scenarios) |
| Human-like self-adaptive ability to unpredictable environments | No | Yes (Abilities vary across scenarios) |
| Initiating capability | Only humans can actively initiate actions, whereas machines passively receive instructions from humans | Both humans and machines can actively initiate actions; AI systems can actively initiate actions based on implicit collaborative interaction (e.g., human behavior, intentions, operational context) |
| Directionality | One-way, human-directed trust and control toward machines without reciprocal situation awareness, trust, and tasks between humans and machines | Two-way and shared trust, situation awareness, and tasks between humans and AI systems (but humans should retain the decision-making authority) |
| Intelligence complementarity | No intelligence complementarity between humans and machines, system optimization mainly depends on the predetermined, static human-machine functional allocation based on human-machine unique strengths | Machine intelligence and human biological intelligence can be dynamically allocated and integrated based on their respective strengths, resulting in hybrid human-machine augmented intelligence. |
| Human intervention | Human intervention is required. Humans must be the ultimate decision-makers | |

As highlighted in Table 1, machines serve as tools to support human operations in human-non-AI system interaction, relying on fixed logic rules or algorithms predetermined by humans, thereby realizing one-way (human-to-machine) human-machine interaction. The machine's behavior is generally passive.

In specific operational environments, AI systems can possess particular cognitive abilities (such as learning, adaptation, and independent execution) enabled by AI technologies, which facilitate the bidirectional "collaborative interaction" between humans and AI systems. For example, this bidirectional "collaborative interaction" in HAII enables AI systems to actively monitor and recognize the user's physiological, cognitive, behavioral, intentional, and emotional states using sensing technology. Meanwhile, human users can obtain the best situational awareness of the system and its environment through multimodal human-machine interfaces. The "collaborative interaction" is driven by the bidirectional, shareable, and complementary characteristics between humans and AI systems (Xu & Ge, 2020; 2021).

It is evident that there are significant differences between these two types of interactions. The new characteristics of AI technologies can enhance the effectiveness of human-machine interaction and facilitate possible human-AI collaborative interactions. However, these new characteristics also bring new challenges different from traditional human-computer interaction, requiring new thinking and approaches to optimize the collaborative interaction between humans and AI systems.

## 2.2 An emerging human-machine relationship brought by AI technologies

Historically, the evolution of human-machine relationships has been driven by technological advancements. As shown in Figure 1, the human-machine relationship evolved from the pre-World War II "human adaptation to machine" paradigm (i.e., the optimized performance of a human-machine system mainly relies on human operator training and skills) to the post-war "machine adaptation to human" paradigm (i.e., the optimized performance of a human-machine system can be achieved through machine design, such as user interface design), and completed the transition from a "machine-centered" approach to a "human-centered" approach. In the PC era, the human-machine relationship continued to evolve into "human-machine interaction", i.e., the optimized performance of a human-machine system relies on the optimal interaction between humans and machines, which can be achieved through human-centered design to ensure the machine design (e.g., user interaction design) matches user needs and user experience. In the "human-machine interaction" paradigm, non-AI computing systems primarily serve as auxiliary tools to support human operations.

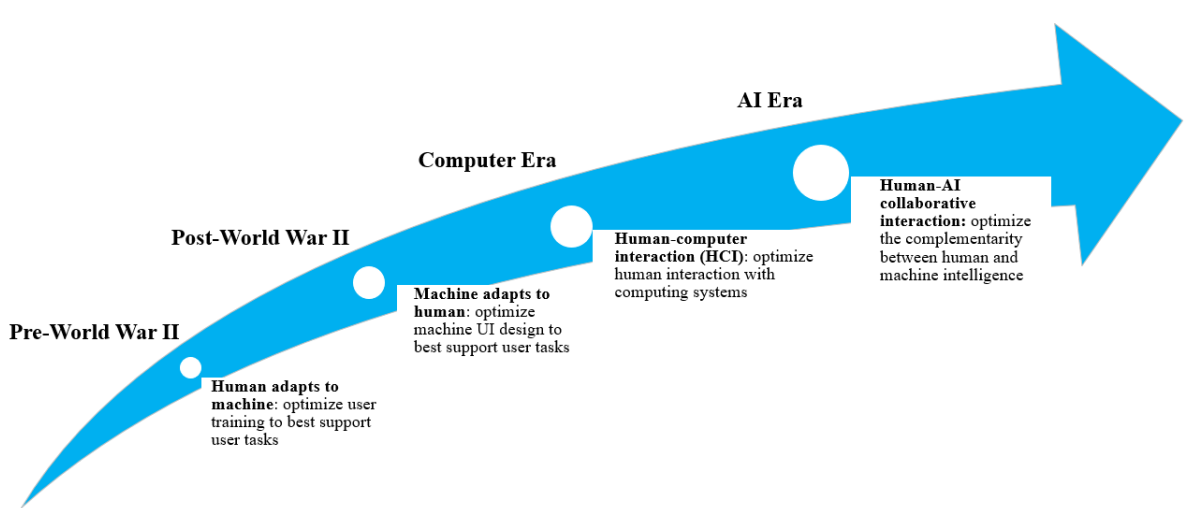

Figure 1 Evolution of human-machine relationships across eras

In the AI era, with the advancement of AI technology, AI systems with autonomous capabilities may evolve from auxiliary tools supporting human operations to collaborative interaction with human operators. AI technology has brought about a new form of human-machine relationship, forming a new type of human-machine relationship, i.e., "human-AI collaborative interaction" (Brill, Cummings, et al.,

2018; Shively, Lachter, et al., 2018; Xu & Ge, 2020; Xu, 2021). Human-AI collaborative interaction refers to the dynamic and synergistic relationship where humans and AI systems work together to achieve shared goals, leveraging each other's strengths and compensating for weaknesses, enhancing overall performance and outcomes

Researchers have raised valid concerns that the new design metaphor of human-AI collaboration (i.e., human-AI teaming) may undermine the HCAI approach (Shneiderman, 2021), potentially misguiding the development of AI systems that lead to a loss of human control during operations. For this very reason, the development of AI systems based on human-AI collaborative interaction must follow the HCAI approach, ensuring that humans retain their leadership roles and final decision-making authority over AI systems (Xu & Gao, 2024; Xu, Gao, Dainoff, 2025).

This new type of human-machine relationship in the AI era signifies that the relationship between humans and machines is not just a tool-like or competitive one, but also a complementary and potentially collaborative one. Currently, human-AI collaboration is becoming a new paradigm for developing AI technology (NAS, 2021). In the "US National AI Research and Strategic Plan 2023 Update", the plan explicitly proposes the need to research human-AI collaboration (NITRD, 2023). Human-AI collaborative interaction not only has characteristics like human-human collaboration, but also has some unique features, which may promote future research on modeling, design, technology, user verification, etc., in aspects such as human-AI bidirectional situation awareness, human-AI mutual trust, human-AI decision-making sharing with human authority, human-AI social interaction, and human-AI emotional interaction.

Although current AI technology falls short of meeting the standard of effective collaborators, the advancement of AI technology will continue to deepen the relationship between humans and AI as their collaborative interaction becomes more sophisticated. In the long run, considering the collaborative interaction between humans and AI will promote the development of AI technology, provided that we design AI systems to retain humans in leadership roles (NAS, 2021; Xu & Gao, 2024). The overall performance of AI systems based on human-AI collaborative interaction depends not only on the performance of individual members within the system, but also on the complementarity between human and machine intelligence. This complementarity can overcome the limitations of each member and maximize the overall system performance.

## 2.3 Transitioning to human-AI interaction (HAII)

With the transition of machines to AI-based intelligent systems in the AI era, a myriad of new characteristics and corresponding research issues have arisen for human-AI interaction (HAII). Table 2 exemplifies some key issues and emerging research needs in HAII (Xu, Dainoff, et al., 2022).

Table 2  Key issues and emerging research needs in human-AI interaction

| Transformative characteristics | New issues from AI technology | Emerging research needs on HAII |
|---|---|---|
| From expected to unexpected machine behavior | • AI systems can bring uncertain machine behavior and unique behavior evolution, leading to potential output bias (Rahwan et al., 2019)<br>• Existing software engineering methods lack consideration of machine behavior<br>• The behavior of AI systems demonstrates characteristics such as evolution and social interaction | • Behavioral science approaches to studying machine behavior<br>• Iterative design and user testing methods to avoid system output bias in data collection, training, and algorithm tuning (Amershi et al., 2014)<br>• User participatory design, human-centered machine learning (Kaluarachchi et al., 2021) |
| From "human-machine interaction" to "human-AI collaborative interaction" | • Machines (AI agents) collaboratively interact with humans<br>• How to model human-AI collaborative interaction (e.g., shared trust, situation awareness, mental models) while retaining humans' leadership roles | • Theories and methods of human-AI collaborative interaction<br>• Design and evaluation approaches of human-AI collaborative interaction (Bansal et al., 2019) |
| From "human intelligence" to "human-machine hybrid enhanced intelligence" | • Machines cannot imitate high-order human cognitive capabilities, and developing machine intelligence in isolation encounters a bottleneck effect (Zheng et al., 2017)<br>• The integration of human roles into AI systems becomes | • Cognitive architecture for human-machine hybrid augmented intelligence<br>• Human-multi-agent collaborative systems based on the human-AI joint cognitive ecosystem paradigm<br>• "Human-in/on-the-loop"-based interaction design |

| | | • Human high-order cognitive models, knowledge representation, and graphs |
|---|---|---|
| | crucial to achieving human-controllable AI (Zanzotto, 2019) | |
| From "human-centered automation" to "human-controllable autonomy" | • Humans may lose ultimate control of AI-based autonomous systems<br>• Potential negative impacts of AI (e.g., indeterministic, biased output)<br>• Confusion between automation and autonomy technologies can lead to an underestimation of the potential negative impacts of AI | • Human-AI interaction design paradigm for autonomous technology<br>• The human factors methods of human-controllable autonomy<br>• Human-machine shared autonomous control design with human authority<br>• Human-autonomy interaction |
| From "non-intelligent" to "intelligent" human-machine interaction | • How to make intelligent user interfaces more natural<br>• How to effectively design human-AI interaction (Google PAIR, 2019)<br>• Bottleneck effect of human perception capability and cognitive resources in a ubiquitous computing environment | • New design paradigms for human-AI interaction and user interface design<br>• Multi-channel natural user interface design<br>• Emerging human-machine interaction technology and design (e.g., emotional interaction, intention recognition, brain-computer interface)<br>• Human-AI interaction design standards for intelligent interaction |
| From "usable" to "explainable AI" design | • The AI "black box" effect can lead to unexplainable and incomprehensible system outputs (Muelle et al., 2019)<br>• AI "black box" effect raises AI trust issues | • Innovative user interface technology (e.g., visualization) and design<br>• "Human-centered" explainable and understandable AI (Ehsan et al., 2021)<br>• Application of explanatory theories in psychology (Mueller et al., 2019) |
| From "user precise input" to "fuzzy reasoning" interaction | • User input is not only a single and precise form (e.g., keyboard, mouse), but also multimodal and fuzzy interaction (e.g., user intention)<br>• The fuzzy interaction-related issues in application scenarios (e.g., random interaction signals and environmental noise) | • Methods and models for inferring user intentions under uncertainty (Yi et al., 2018)<br>• The naturalness and effectiveness of human-AI interaction under fuzzy conditions |
| From "user experience" to "ethical AI" | • New user needs (privacy, ethics, fairness, reskilling, decision-making authority, etc.) (IEEE, 2019)<br>• Potential output bias and unexpected results of AI systems<br>• Abuse of AI systems (discrimination, privacy, etc.)<br>• Lack of traceability and accountability mechanisms for AI system failures | • Multidisciplinary design for ethical AI<br>• Sociotechnical approach of ethical AI<br>• Meaningful human control (Santoni & van den Hoven, 2018)<br>• Transparency design |
| From "experience-based" to "systematic" interaction design | • Limitations of design methods in current user experience (UX) practices<br>• UX design challenges of AI systems<br>• Human factors science professionals failed to intervene early in the development of AI systems in many cases | • Human-centered process for developing AI systems<br>• AI-based innovative design driven by user experience<br>• Effective intelligent interaction design methods (Holmquist, 2017)<br>• Systematic human-centered methods (Xu et al., 2019) |

As discussed earlier, AI technology introduces many new characteristics to HAII, as well as a new type of human-machine relationship. AI technology and HAII not only bring great potential value to humans but also bring a series of new challenges. These new challenges have far exceeded the scope of existing research (e.g., human-computer interaction) that focuses on human interactions with non-AI computing systems, necessitating design thinking and approaches.

These new challenges also present opportunities to explore innovative approaches, thereby developing human-centered AI systems that enable AI technology to benefit humans to the greatest extent and minimize harm to them. The emergence of human-AI interaction (HAII), systematically promoted in this book, is one way to achieve this goal.

Currently, the vast majority of AI technology research and development is focused on developing human-facing AI systems, such as intelligent user interfaces, intelligent robots, automated driving vehicles, intelligent decision-making systems, generative AI, and intelligent healthcare solutions. Humans directly interact with these AI systems through user interfaces, processes, and services, and the effectiveness of these interactions directly relates to important issues involving the HCAI approach, such as how AI systems meet human needs and values, how AI technology serves humans and society, and how to mitigate adverse impacts on humans.

Research addressing these issues has been initiated, including studies on "Human-Intelligent System Interaction" (Brill et al., 2018), "Human-Agent Interaction" (Salehi et al., 2018), "Human-Autonomy Interaction" (Cummings & Clare, 2015), and "Human-AI Interaction" (Amershi et al., 2019; Xu, Ge, & Gao, 2021). Although these studies have distinct emphases, they collectively explore interactions between humans and AI-based intelligent machines, that is, human-AI interaction (HAII).

### 3 A human-centered perspective on human-AI interaction (HAII)
### 3.1 The need for a human-centered perspective on HAII
### 3.1.1 Human-centered AI approach

AI technologies have significantly benefited humans and society, but they have also caused harm when poorly designed, developed, or used. Research has identified AI limitations, such as vulnerability, bias, lack of explainability, absence of causal models, and ethical concerns (NAS, 20221; Endsley, 2023). Databases like the AI Accident Database and the AIAAIC database have recorded thousands of AI-related accidents and incidents (McGregor, 2025; AIAAIC, 2025).

A human-centered AI (HCAI) approach has been proposed to address the limitations of traditional AI methods (Shneiderman, 2020, 20222; Xu, 2019; Capel et al., 2023; Schmager et al., 2025). Researchers such as Shneiderman (2020) and Xu (2019) have initially introduced their systematic human-centered AI (HCAI) frameworks, focusing on human values, needs, knowledge, and roles throughout the design, development, deployment, and use of AI systems. HCAI aims to create AI systems that amplify, augment, and enhance human performance in ways that make AI systems reliable, safe, and trustworthy, instead of replacing humans (Shneiderman, 2020; Xu, 2019). This approach has been further developed in areas like humanistic AI design (Auernhammer, 2020), trustworthy AI (Wickramasinghe et al, 2020), human-centered machine learning (Wortman et al., 2021), and human-centered explainable AI (Rong et al., 2024).

More specifically, Shneiderman (2020, 2022) defines HCAI as a design philosophy and development strategy focused on creating AI systems that are reliable, safe, and trustworthy. Rather than seeking to replace humans, HCAI systems aim to augment human capabilities and ensure meaningful human oversight throughout the AI lifecycle. Furthermore, Shneiderman (2020) introduced a two-dimensional framework for HCAI, characterized by the degree of human control and the degree of AI autonomy (see Figure 2). This framework emphasizes that greater autonomy does not necessarily mean less human control; instead, it encourages the design of AI systems that must be controllable by humans.

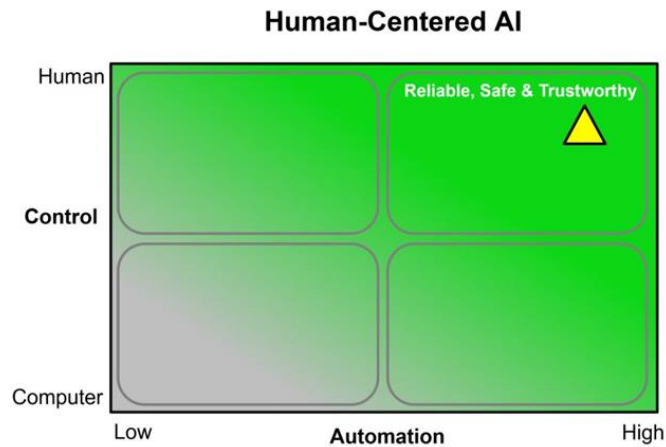

Figure 2  The two-dimensional framework with the goal of reliable, safe, and trustworthy AI

Xu (2019) proposed a "human factors-technology-ethical alignment" HCAI framework that integrates three key perspectives: human factors, technology, and ethics (see Figure 3). From a technology perspective, HCAI combines human and machine intelligence to enhance human capabilities with advanced AI technologies. The human factors perspective emphasizes the need for AI systems to be controllable, user-friendly, and explainable. The ethics perspective addresses critical issues, including human values, fairness, and privacy. HCAI emphasizes a systematic design approach that fosters synergy and complementarity among these perspectives. An HCAI solution achieves balance across all three aspects. To achieve its goals, HCAI underscores the importance of interdisciplinary collaboration, ensuring a holistic approach to HCAI practice.

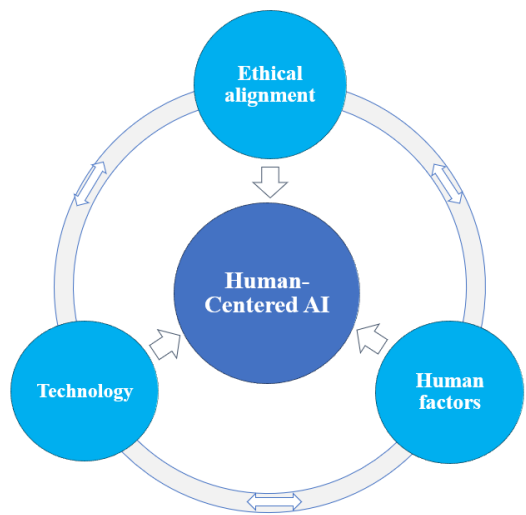

Figure 3  The "human factors-technology-ethical alignment" HCAI framework

### 3.1.2 Challenges in current HAII practices

On one hand, many user-facing HAII-related projects failed, often due to insufficient focus on "human factors" such as poor user experience, lack of explainability and controllability, unmet user needs, and neglect of ethical considerations (Yampolskiy, 2019). AI professionals often claimed that many problems that traditional HCI practices could not solve in the past have been addressed through AI technology (e.g., voice input), allowing them to design interactions independently (Yang et al., 2018). However, while some AI-enabled designs are innovative, studies have shown that their outcomes may fall short from a human-centered perspective (e.g., Budiu & Laubheimer, 2018). This is further evidenced by documented cases in the two AI incident databases.

On the other hand, research indicates that professionals in the human-centered fields, such as human factors, HCI, and user experience, face challenges in integrating human-centered approaches into the development of AI systems. They struggle to collaborate effectively with AI professionals due to a lack of a shared workflow and common language (Girardin & Lathia, 2017). Studies have shown that these professionals do not appear to be adequately prepared to provide practical design support for AI systems (Yang et al., 2018). More specifically, several challenges have been identified by studies as follows.

*Process Challenges:* HAII practices frequently encounter systemic process-level challenges that hinder their integration into AI development lifecycles. Many human-centered activities are introduced too late in the design process, after critical technical decisions have already been made (Silberg et al., 2019). Early AI design continues to rely on traditional software development methods, limiting the influence of human-centered practitioners (Hartikainen et al., 2022). Additionally, a human-centered approach is often separated from technical work, with business clients rather than end-users dictating requirements (Bingley et al., 2023; Mazarakis et al., 2023). As a result, important HCAI principles, such as usability, ethics, and user empowerment, are inconsistently applied or lost altogether (Silberg et al., 2019). These misalignments lead to poor user experiences and a lack of explainability and controllability in AI systems.

*Design Paradigm Challenges:* Current HAII practices tend to focus narrowly on individual human-AI interactions, lacking a comprehensive design paradigm that addresses broader sociotechnical and ecosystem-level concerns (NAS, 2021; Taddeo et al., 2023). This isolated approach misses the opportunity to consider AI's impact within larger organizational, social, or cultural contexts. The absence of system-level thinking and diverse design frameworks limits innovation and reduces the scalability and applicability of HAII solutions. Consequently, human-centered design in AI often remains fragmented and reactive rather than proactive and strategic.

*Method Challenges*: The practical implementation of HAII is hindered by a shortage of actionable and practical human-centered design methods (van, 2018; Umbrello et al., 2024). Professionals from HCI or human factors backgrounds often encounter difficulties in generating diverse AI interaction concepts, creating meaningful prototypes, and conducting robust testing of AI behaviors. Current methodologies are not well-suited for the unique characteristics of AI systems, such as their unpredictability and learning capabilities, which necessitate new approaches. There is a growing need to evolve and adapt human-centered methods to be more practical, iterative, and supportive of the dynamic nature of AI development.

*Interdisciplinary Collaboration Challenges*: Interdisciplinary collaboration, a crucial component of HAII, also faces several challenges (Mazarakis et al., 2023; Yang et al., 2020; Zeller et al., 2022). These include conflicting methodologies, insufficient user involvement, differing terminologies across disciplines, and resistance to collaboration. Team structures are often fragmented, making it difficult to coordinate across technical, design, and social domains. Moreover, scaling such interdisciplinary efforts poses a challenge in large projects or organizations. Without effective collaboration and mutual understanding, it becomes challenging to integrate the diverse perspectives required for delivering HCAI systems.

In summary, these challenges underline the importance of adopting a human-centered approach to HAII practices.

### 3.1.3 HCAI guiding principles and their implications for HAII

The human-centered perspective views HAII not merely as a technical AI-based user interface, but as encompassing all aspects of human interaction with AI systems, from micro interactions to a macro sociotechnical perspective that prioritizes human values, needs, capabilities, and well-being when humans interact with AI systems. The human-centered perspective provides a foundational lens for analyzing human interaction with AI systems through human-centered principles, as elaborated in the HCAI guiding principles that aim to achieve HCAI design goals based on previous studies (see Table 3 – adapted from Xu, Gao, Dainoff, 2025) (Amershi et al, 2019; Google PAIR, , 2019; Floridi et al., 2018; Richards et al., 2023).

Table 3   HCAI Guiding Principles

| # | HCAI Guiding Principles | Definition and HCAI Goals |
|---|---|---|
| 1 | Transparency and Explainability | Provide explainable and understandable AI output for humans to enhance human trust and empower informed decisions. |
| 2 | Human Control and Empowerment | Allow humans to understand, influence, and control AI behavior when necessary, retain human authority and controllability over AI systems instead of replacing humans |
| 3 | Human Value and Ethical Alignment | Develop AI in alignment with ethical and societal norms to preserve human values and privacy, foster trust, and minimize harm. |
| 4 | User Experience | Create interactions that are engaging, intuitive, accessible, and aligned with user needs and expectations. |
| 5 | Human Augmentation and human-les collaborative interaction with AI | Design AI to enhance human abilities and facilitate human-led collaborative interaction, thereby improving productivity and effectiveness. |
| 6 | Safety and Robustness | Prioritize human safety, maintain reliability in diverse scenarios to ensure resilience and reduce potential risks to humans. |
| 7 | Accountability and Responsibility | Ensure responsible AI mechanisms for accountability to hold humans (e.g., developers, operators, commanders) accountable for AI actions. |
| 8 | Sustainability | Develop AI to support environmental, social, and economic well-being to prioritize human well-being while aligning with sustainability. |

*Transparency and Explainability*. Transparent and explainable AI systems allow humans to comprehend why the system made a particular decision or recommendation. This is especially important in high-stakes domains (e.g., healthcare, finance) where interpretability influences trust, acceptance, and oversight. AI systems should clearly communicate how decisions are made, enabling users to understand system logic, foster trust, and support informed interactions. User interfaces must visualize reasoning and offer traceable outcomes. HAII research and applications must develop user-friendly visualization, natural language explanations, and interaction mechanisms to reveal system rationale without overwhelming users.

*Human Control and Empowerment*. AI systems must enable graduated levels of human involvement, ranging from human-in-the-loop to human-on-the-loop, to accommodate varying levels of autonomy and control. AI systems must allow humans to intervene, override, or modify AI behavior when necessary, retaining human authority and controllability over AI systems. Interaction paradigms should empower humans rather than automate away critical decision-making roles. HAII research and applications should design user interfaces and feedback mechanisms that allow humans to override, retrain, or redirect AI behavior. This supports not only technical control but also psychological empowerment, reducing feelings of helplessness and increasing human agency, particularly in shared autonomy or assistive AI systems.

*Human Value and Ethical Alignment*. Human values and ethical alignment design require an understanding of human values, contextual norms, and cultural variations in moral expectations. AI systems should respect human values, social norms, and cultural expectations. For example, ethical design choices must be embedded into conversational agents, decision support tools, and feedback mechanisms. HAII research and applications must integrate ethical reasoning into system behavior and design interaction paradigms, aligning with personal, organizational, and social norms. This principle also calls for value-sensitive design and human-centered participatory methods that involve stakeholders and end-users in shaping AI decision logic, particularly in socially sensitive applications (e.g., hiring, surveillance, education).

*User Experience.* User experience (UX) extends beyond usability to encompass emotional, cognitive, and social dynamics of human interaction with AI systems. HAII research and applications should emphasize interaction design that adapts to diverse users, minimizes friction, and builds sustained engagement. This includes multimodal user interfaces, emotional responsiveness, and adaptive personalization. By aligning with user mental models and expectations, AI systems can reduce human error, increase efficiency, and improve satisfaction.

*Human Augmentation & Human-Led Collaborative Interaction with AI.* Interaction design should focus on enhancing human capabilities through shared tasks, decision support, or creative augmentation, rather than simply replacing human roles. Design must retain humans' authority and leadership role in possible human-AI collaboration. AI systems should aim to complement human strengths with AI capabilities. This principle guides the design of collaborative user interfaces where roles, goals, responsibilities, and human authority are clearly defined. HAII research and applications should investigate team dynamics, shared mental models, adaptive roles, and human authority in human-AI collaborative interaction, particularly in complex tasks like mission planning, co-driving, or medical diagnostics.

*Safety and Robustness.* AI systems should exhibit safe and predictable behavior even in uncertain situations. Interaction modalities should help humans detect anomalies, recover from failures, and maintain situation awareness. Safety-critical applications require AI systems to behave predictably under uncertainty and maintain functionality during disruptions or adversarial conditions. HAII research and applications must incorporate features such as fail-safe interaction design and real-time anomaly detection to preserve safety and prevent harm. This includes user alerts, emergency shutdowns, and transparent recovery processes in contexts such as autonomous vehicles, industrial robotics, or healthcare devices.

*Accountability and Responsibility.* AI systems should clearly indicate who is responsible for their outputs and actions. Interactions should support traceable interaction logs, decision audit trails, and clear attribution of responsibility among human and AI agents. This supports both retrospective analysis and forward-looking governance, enabling users, regulators, and stakeholders to understand who is responsible for outcomes. HAII research and applications must address user interface-level accountability cues and human oversight models in AI systems.

*Sustainability.* AI systems should be designed with technical, environmental, economic, and social sustainability in mind. HAII research and applications should promote responsible AI usage, reduce unnecessary resource consumption (e.g., excessive data processing or energy use), and support the long-term well-being of humans and society. Achieving sustainability also involves adopting integrative approaches, such as data and knowledge dual-driven AI as well as human-AI hybrid enhanced intelligence, that help avoid the siloed technical approach to AI development. These approaches are critical for overcoming AI development bottlenecks and ensuring that AI evolves in a scalable, adaptable, and sustainable manner.

### 3.2 A conceptual framework of human-centered HAII (HC-HAII)

Although HAII-related research and applications are growing, there is currently no mature, systematic framework for HAII that encompasses research objectives, design philosophy, scope, and methodology from the HCAI perspective (Xu, Ge, & Gao, 2021). Driven by the HCAI approach, this chapter proposes a conceptual framework of human-centered HAII (HC-HAII) (see Figure 4). The HC-HAII framework reflects the following characteristics.

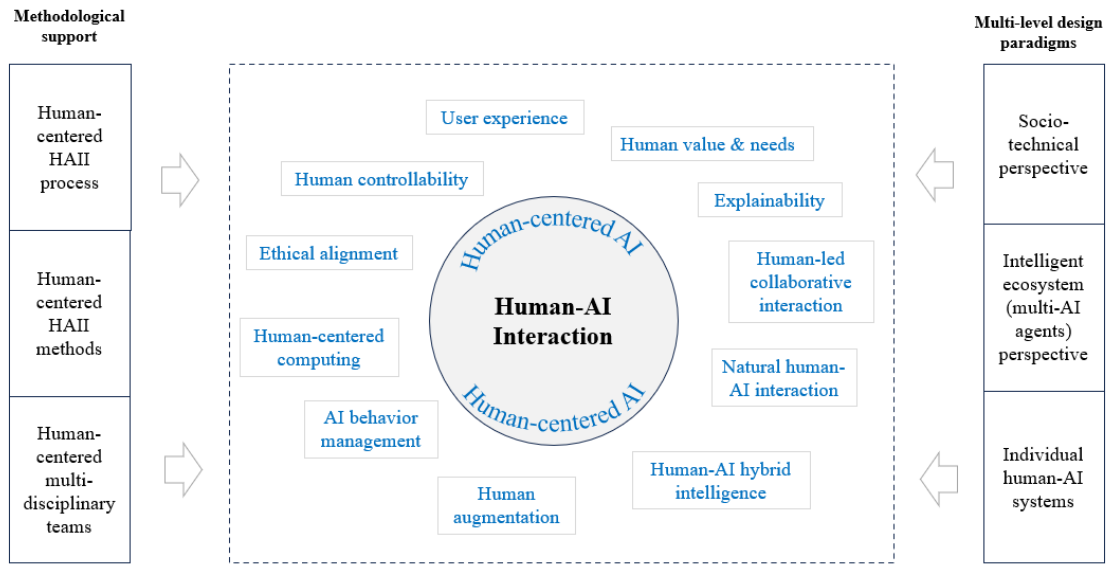

Figure 4  A conceptual framework of human-centered human-AI interaction (HC-HAII)

- *Aims*. HC-HAII emphasizes the adoption of the HCAI approach in HAII research and applications. It is dedicated to researching and developing HCAI systems in alignment with HCAI guiding principles. HC-HAII focuses on the new challenges brought by AI technology, emphasizing the complementary advantages of human intelligence and machine intelligence. It is committed to optimizing the interaction between humans and AI systems, fully considers human needs and AI ethics, and emphasizes that humans have the final decision-making control over AI systems. HC-HAII aims to implement the HCAI design philosophy in the design, development, deployment, and use of AI systems through HAII research and applications, promoting and developing beneficial AI systems for humans while minimizing the negative impact of AI on them.

- *Scope and key research issues*. HC-HAII is dedicated to addressing the critical challenges in designing, developing, deploying, and using user-facing AI systems from the HCAI perspective. The representative issues are outlined in the HC-HAII framework (see blue text boxes in Figure 4) and reflect the core requirements derived from the HCAI guiding principles. Effectively addressing these challenges is essential for advancing HAII research and applications, enabling the development of truly human-centered AI systems. See Section 3.4 for further details.

- *Methodology*. HC-HAII fosters interdisciplinary collaboration and integrated innovation. Building on this foundation, HC-HAII places a stronger emphasis on effective collaboration, particularly with human-centered disciplines such as human factors, human-computer interaction (HCI), and user experience (UX). HC-HAII advances human-centered methodologies, encompassing specialized human-centered methods, processes, multidisciplinary collaboration, and multi-level design paradigms as illustrated in the left-hand and right-hand columns of Figure 4. These approaches are essential for operationalizing the HCAI guiding principles and achieving HCAI design goals. The interdisciplinary methods represented in this framework are drawn from diverse fields, including AI, computer science, data science, HCI, human factors engineering, cognitive neuroscience, engineering psychology, cognitive science, social sciences, law, and ethics. The contributing authors of this book represent more than ten academic and professional disciplines, highlighting the integrative nature of the HC-HAII approach. See Section 4.0 for details.

- *Design paradigms*. A design paradigm offers a specific lens that shapes the perspective, scope, approaches, and methods of a field. HC-HAII promotes multi-level design paradigms, each grounded in the HCAI perspective as illustrated in the right-hand column of Figure 4. Narrowly defined, HC-HAII addresses issues arising from direct interaction between humans and individual AI systems. Broadly, it extends HAII research and applications to encompass the broader AI ecosystem and sociotechnical contexts, going beyond isolated human-AI interactions. HC-HAII emphasizes that HAII research and applications should account for the influence of diverse factors on human-AI interaction, including ecosystem-level considerations (e.g., multi-agent systems, human-AI teams, AI-AI coordination) and sociotechnical dimensions (e.g., organizational, cultural, and ethical systems). The content of this book reflects the macro, integrative perspectives, enabling HAII to deliver comprehensive, human-centered solutions. See Section 4.4 for details.

### 3.3 Why human-centeredness matters in HAII

Globally, HAII research and applications are still in their early stages. This book seeks to advance the field by providing the following contributions from the HCAI perspective.

Firstly, HC-HAII underscores that its research and applications are centered on AI systems, in contrast to the traditional field of human-computer interaction, which primarily focuses on non-AI computing systems. Promoting HC-HAII draws critical attention to the fundamental differences between AI and non-AI systems, urging researchers and practitioners to recognize the unique challenges introduced by AI technologies. It encourages a shift away from legacy mindsets and promotes the adoption of innovative approaches tailored to the complexities of human-AI systems. Historically, technological advances have catalyzed the emergence of new research fields. For example, the advent of PC in the 1980s marked a shift from "human-machine interaction" to the then-new field of "human-computer interaction." Today, the transition from traditional non-AI computing systems to AI-based intelligent systems signals the rise of the AI era. The distinct characteristics of AI technologies have necessitated the formation of the HAII field. At the same time, HC-HAII ensures that HAII research and applications effectively address the challenges identified in current HAII practices (see Section 3.1.2) and remain committed to delivering human-centered systems.

Secondly, promoting HC-HAII supports the advancement and practical implementation of HCAI in HAII research and applications. In this means–end relationship, HCAI represents the overarching goal, while HAII serves as a critical pathway to achieving it, each reinforcing the other. This emphasis enhances understanding of the HCAI approach and encourages the adoption of effective human-centered methods for realizing its objectives. By leveraging the complementary strengths of human and machine intelligence, HC-HAII advances interaction design, prioritizes ethical considerations, and integrates human centrality into AI systems to mitigate safety risks. Moreover, it reinforces the principle that AI should be designed to augment humans, not to replace them.

Finally, HC-HAII fosters cross-disciplinary and cross-industry engagement, particularly involving key fields such as human-computer interaction and human factors. This collaboration promotes integrated efforts among professionals from diverse domains, enabling the co-development of human-centered AI systems within a unified HAII framework. The interdisciplinary author team of this book exemplifies such collaboration, which also contributes to the advancement of interdisciplinary methods for implementing HCAI systems.

### 3.4  Emerging themes in HC-HAII

Every field has its fundamental research and application themes for study. As an emerging interdisciplinary field, HAII is developing its foundational themes to address. Table 4 presents some challenges faced in current HAII practice and HC-HAII solutions covered in this book across emerging fundamental themes. These HC-HAII solutions are grounded in the HCAI approach, offering a foundation for advancing HAII research and applications from the HCAI perspective. The scope of HC-HAII research and applications is broad, and the content listed in Table 4 represents only a portion of it. With the advances in AI technology, new themes will continue to emerge, and the fundamental research and application themes of HC-HAII need to be continuously improved and expanded in future practice.

Table 4  Challenges faced in current HAII practice and possible HC-HAII solutions

| Emerging themes (examples) | Challenges faced in implementing HAII | HC-HAII solutions (examples) | Primary HCAI guiding principles applied | Related chapters of this book |
|---|---|---|---|---|
| AI machine behavior | • AI systems can learn autonomously, bringing about issues such as output uncertainty, algorithm bias, unexpected behavior, and evolving behavior<br>• Existing machine learning training and testing methods lack active human participation<br>• There are complex machine behaviors and interactions among multiple AI agents | • Human-centered, interactive, user-participatory machine learning<br>• Human-centered algorithm nudge, intelligent recommender systems<br>• Human-centered methods in data collection, algorithm training and tuning<br>• Machine behavior design and management based on behavioral science | • Transparency and Explainability<br>• Human Value and Ethical Alignment<br>• Safety and Robustness | • Human-centered explainable AI (Chapter 13)<br>• Designing Nudges and Trust: Toward Ethical and Human-Centered AI (Chapter 15)<br>• Machine behavior design and management (Chapter 20) |
| Human modeling | • Machine intelligence cannot emulate higher-order cognitive capabilities of humans<br>• AI relying solely on big data has problems such as poor reasoning, strong dependence on big data and computing power, and unexplainability<br>• There is a lack of effective human state recognition models (user intent, situational awareness, emotions, etc.) | • Human cognition and behavior modeling<br>• Human knowledge representation and knowledge graphing<br>• Human state modeling (cognitive, physiological, emotional, intentional, etc.)<br>• Human performance modeling | • Human Value and Ethical Alignment<br>• Human Augmentation and human-les collaborative interaction with AI<br>• User Experience | • Computational modeling of human cognition and behavior (Chapter 2)<br>• Human-centered human-AI collaboration (HCHAC) (Chapter 9)<br>• Human state recognition in HAII (Chapter 12) |
| Human-centered computing | • Unexplainable algorithms<br>• Model bias and unfairness<br>• Limited user control or feedback mechanisms<br>• Misalignment with individual or societal values<br>• Misalignment with user goals<br>• Lack of user trust and understanding | • Human-centered machine learning<br>• Human-centered collaborative computing<br>• Human-centered explainable AI<br>• Human-centered social computing<br>• Human-centered recommender systems<br>• Human-centered algorithmic nudge | • Human Value and Ethical Alignment<br>• Human Augmentation and human-les collaborative interaction with AI<br>• Human Control and Empowerment | • Human-centered explainable AI (Chapter 13)<br>• Human-centered machine learning (Chapter 14)<br>• Designing Nudges and Trust: Toward Ethical and Human-Centered AI (Chapter 15)<br>• Recommender systems: A human-centered AI perspective (Chapter 16)<br>• Social computing: A human-centered AI perspective (Chapter 17) |
| Human intelligence augmentation (IA) | • Lack of consideration of the complementary advantages of AI and IA.<br>• Machine intelligence cannot mimic specific dimensions of human intelligence, requiring interdisciplinary exploration of IA technologies and methods.<br>• Lack of practical tools and methods based on the neuroplasticity mechanism of human brain cognition.<br>• Lack of effective interaction-based technologies (e.g., cognitive disease diagnosis and rehabilitation) | • Human-centered IA and AI-based human IA technologies (e.g., cognitive enhancement).<br>• Human IA technologies and methods based on plasticity mechanisms, controllable cognitive load, physiological feedback, and body-brain interaction.<br>• Integration schemes of AI and IA technologies at the neural level (e.g., brain-computer fusion) | • Human Augmentation and human-les collaborative interaction with AI<br>• Human Control and Empowerment | • AI-based human intelligence enhancement (Chapter 5)<br>• Brain-computer interface and brain-computer fusion (Chapter 8)<br>• Human-AI interaction in assistive technology (Chapter 12)<br>• Human-centered approach for AI-based rehabilitation technology and accessibility design (Chapter 21).<br>• Patient-centered intelligent healthcare (Chapter 23).<br>• Child-centered AI-assisted autism diagnosis and treatment (Chapter 24) |
| Human-AI hybrid | • Machine intelligence has difficulty modeling human higher cognitive | • "Human-in-the-loop" human-machine hybrid enhanced intelligence | • Human Control and Empowerment | • Data and Knowledge Duel-Driven AI (Chapter 4) |
| enhanced intelligence | abilities; the isolated development of AI lacks sustainability.<br>• Lack of consideration for "human-in/on-the-loop" in AI systems leads to safety issues (e.g., human control)<br>• Human intelligence has not achieved the maximum degree of complementary advantages in cognition and behavior<br>• AI technology solely dependent on big data lacks sustainability | • "Human-on-the-loop" human-machine hybrid enhanced intelligence<br>• Brain-machine fusion<br>• Data and knowledge dual-driven AI | • Safety and Robustness<br>• Sustainability | • Brain-computer interface and brain-computer fusion (Chapter 8)<br>• Recommender systems: A human-centered AI perspective (Chapter 16)<br>• Human-centered approach for AI-based rehabilitation technology and accessibility design (Chapter 21) |
| Human-led collaborative interaction | • Existing human-machine interaction is based on the "machine as a tool" paradigm, lacking consideration of potential human-AI collaborative interaction<br>• Lack of mature theories, methods, cognitive architectures, models, and evaluation methods for human-AI collaborative interaction<br>• Lack of mature theories, models, and methods for shared situation awareness, trust, and mental models | • New design paradigm of human-AI collaborative interaction<br>• New theories, models, cognitive architecture, evaluation, and prediction methods for human-AI collaborative interaction<br>• Interaction models of human-AI collaborative interaction<br>• Human-AI collaborative user interface | • Human Augmentation and human-les collaborative interaction with AI<br>• User Experience | • Human-centered human-AI collaboration (HCHAC) (Chapter 9)<br>• Interaction models and design paradigms of human-AI interaction (Chapter 19)<br>• Human-centered collaborative automated driving (Chapter 22) |
| Explainable AI | • The "black box" effect of AI makes users unable to understand AI decision output, affecting human trust<br>• Psychological explanation theory has not been fully utilized<br>• Lack of effective user-participatory, explainable AI methods<br>• Many explainable AI projects are adopting an "algorithm-centered" method, exacerbating the problem of algorithm opacity | • Human-centered explainable AI<br>• User-participatory explainable AI<br>• "Human-in/on-the-loop" style explainable AI<br>• Application transformation of psychological explanation theory<br>• Natural, interactive AI human-machine interface design | • Transparency and Explainability<br>• User Experience | • Human-centered explainable AI (Chapter 13)<br>• Designing Nudges and Trust: Toward Ethical and Human-Centered AI (Chapter 15) |
| Human-controllability | • Human factors issues in system design, such as decline in operator situation awareness, mode confusion, human-out-of-the-loop, and over-trust<br>• The confusion between automation and autonomy, underestimating the impact of autonomous technology | • Meaningful human control<br>• Human-controllable autonomy<br>• Human-machine shared autonomy (such as takeover/handover in emergencies)<br>• Human-in/on-the-loop design in human-AI interaction | • Human Control and Empowerment<br>• Safety and Robustness | • Meaningful human control in AI (Chapter 7)<br>• Human-centered collaborative automated driving (Chapter 22)<br>• Pilot-centered intelligent civil aircraft cockpit (Chapter 26) |
| Nature human-computer interaction | • Lack of mature theories and technologies for intelligent human-computer interaction, such as HAII models, social interaction, emotional interaction, and user intent recognition | • Physiological, emotional, and intent modeling in intelligent human-computer interaction<br>• Theories, models, and technologies of intelligent human-computer interaction | • User Experience<br>• Human Control and Empowerment | • Human state recognition in human-AI interaction (Chapter 3)<br>• Human factors in human-AI interaction (Chapter 6)<br>• Multimodal interaction with wearable computing devices (Chapter 10) |
| | • Lack of effective interaction models and design paradigms for AI systems<br>• Bottleneck effect of limited human cognitive resources in complex intelligent computing environments<br>• Lack of mature methods for experience design in HAII<br>• Lack of mature HAII design standards and guidelines | • New paradigms of intelligent human-computer interaction<br>• Design standards for intelligent human-computer interaction<br>• User experience design of human-AI interaction | | • Human-AI interaction in the metaverse and virtual reality (Chapter 11)<br>• Human AI interaction design standards (Chapter 18)<br>• Interaction models and design paradigms of human-AI interaction (Chapter 19) |
| Ethically aligned design | • AI systems may produce output bias and unexpected results<br>• AI abuse issues (discrimination, privacy leaks, etc.)<br>• Lack of human ultimate control over AI systems<br>• Lack of traceability and accountability mechanisms for AI system failures<br>• Human-centered AI ethical design | • Algorithm governance<br>• Meaningful human control, AI accountability mechanism<br>• Transparent system design<br>• Iterative prototype design and user testing | • Human Value and Ethical Alignment<br>• Accountability and Responsibility<br>• Transparency and Explainability | • Meaningful human control in AI (Chapter 7)<br>• Human-centered explainable AI (Chapter 13)<br>• Designing Nudges and Trust: Toward Ethical and Human-Centered AI (Chapter 15)<br>• Machine behavior design and management (Chapter 20) |

According to Table 4, the realization of the HCAI approach presents a series of challenges in HAII practice. How to solve these new challenges directly relates to whether the HCAI approach, as driven by HC-HAII, can be realized. Currently, research on these fundamental themes of HC-HAII is underway. Table 4 also looks forward to HC-HAII solutions (see column 3). For detailed content, please refer to the relevant chapters of this book.

## 4 HC-HAII methodology

The HC-HAII framework is not an abstract concept; it includes its methodology, including human-centered methods, process, multidisciplinary collaboration, and multi-level design paradigms.

### s4.1 Human-centered methods

In response to the new characteristics and challenges of AI technology, HC-HAII inevitably requires effective methods. The interdisciplinary nature of the HC-HAII field determines its methodology. Table 5 summarizes the interdisciplinary methods presented in this book. These interdisciplinary methods are based on the HCAI approach, which places human needs, values, wisdom, abilities, and roles at the forefront of AI design, development, deployment, and use. Many of these methods come from non-AI disciplines or are enhancements of existing AI discipline methods based on the HCAI design philosophy, such as human-centered explainable AI, human-centered recommender systems, and human-centered social computing. These methods complement each other and contribute to the development of HCAI systems.

Table 5   Human-centered methods in HC-HAII

| Category | Methods (examples) | Related chapters in this book |
|---|---|---|
| Overall approaches | • Human-centered human-AI interaction<br>• Brian-computer interface and fusion<br>• Data and knowledge dual-driven AI<br>• Human intelligence enhancement<br>• AI machine behavior management | • Human-centered human-AI interaction (HC-HAII): A human-centered AI perspective (Chapter 1)<br>• Data and Knowledge Dual-Driven AI (Chapter 4)<br>• AI-Based Human Intelligence Enhancement (Chapter 5)<br>• Brain-Computer Interface and Brain-Computer Fusion (Chapter 8)<br>• Machine Behavior Design and Management (Chapter 20) |
| Research Paradigms | • Human-AI joint cognitive systems<br>• Human-AI joint cognitive ecosystems<br>• Intelligent sociotechnical systems | • Human-centered human-AI interaction (HC-HAII): A human-centered AI perspective (Chapter 1)<br>• Human-centered human-AI collaboration (HCHAC) (Chapter 9)<br>• Interaction models and design paradigms of human-AI interaction (Chapter 19) |
| Models (theory, computation) | • Human state (physiological, emotional, intentional, etc.)<br>• Human cognition and behavior modeling<br>• Human performance modeling<br>• Human-AI interaction models | • Computational modeling of human cognition and behavior (Chapter 2)<br>• Human state recognition in human-AI interaction (Chapter 3)<br>• Interaction models and design paradigms of human-AI interaction (Chapter 19) |
| Algorithms | • Human-centered perceptual computing<br>• Multi-channel human-AI collaborative computing<br>• Human-centered recommender systems<br>• Human-centered explainable AI<br>• Human-centered algorithm nudge | • AI-Based Human Intelligence Enhancement (Chapter 5)<br>• Brain-Machine Interface and Brain-Machine Fusion (Chapter 8)<br>• Human-Centered Explainable AI (Chapter 13) |
| | • Human-centered social computing<br>• Brain-computer interface and fusion<br>• Human intelligence enhancement | • Designing Nudges and Trust: Toward Ethical and Human-Centered AI (Chapter 15)<br>• Recommender systems: A human-centered AI perspective (Chapter 16)<br>• Social computing: A human-centered AI perspective (Chapter 17) |
| User Research, Interaction Design, User Validation | • Human-centered AI design<br>• Interaction design<br>• User experience design and validation<br>• Human factors methods of HAII<br>• New design paradigms of HAII<br>• Human-AI interaction design standards | • Human factors in human-AI interaction (Chapter 6)<br>• Multimodal Interaction with Wearable Computing Devices (Chapter 10)<br>• Human-AI interaction in the metaverse and virtual reality (Chapter 11)<br>• Human AI interaction design standards (Chapter 18)<br>• Machine Behavior Design and Management (Chapter 20) |
| Standards and Governance | • Ethical AI standards and governance<br>• HAII design standards<br>• Responsible AI through Meaningful human control | • Meaningful Human Control in AI (Chapter 7)<br>• Human AI interaction design standards (Chapter 18) |

The interdisciplinary methods outlined in Table 5 encompass the entire AI lifecycle, encompassing overall approaches, research paradigms, models (both theoretical and computational), algorithms, interaction technologies and design, user research and validation, standards development, and AI governance. As an emerging field, these methods are not yet fully mature and require ongoing enrichment and improvement.

## 4.2 Human-centered process

Any field requires an effective implementation process to realize the design goals of the field. Based on the HCAI approach, HC-HAII adopts a human-centered process spanning the entire AI product lifecycle, prioritizing human needs, values, capabilities, and roles at the forefront of AI system design, development, deployment, and use. This means integrating HCAI-based methods and activities in the end-to-end process across the AI lifecycle. The "end-to-end" process means that HC-HAII work is not limited to the design and development stages of AI systems, but also includes stages after development, such as deployment and use. For example, this includes ethical AI governance, addressing user privacy concerns, fairness issues, and monitoring the behavior of AI machines. Through such an end-to-end process, the HCAI approach can be fully and effectively implemented.

At present, there is a lack of a comprehensive process to support HAII work. As shown in Figure 5, the HC-HAII process framework adopts the human-centered design (HCD) process defined by the International Organization for Standardization (ISO) (ISO, 2019). The HCD process is a human-centered, iterative approach focused on addressing human needs and experiences. It places the user at the forefront of the system lifecycle stages, including design, development, deployment, and use, ensuring that solutions meet user needs through iterative design methods that involve collecting and defining user needs, interaction prototyping, and user validation.

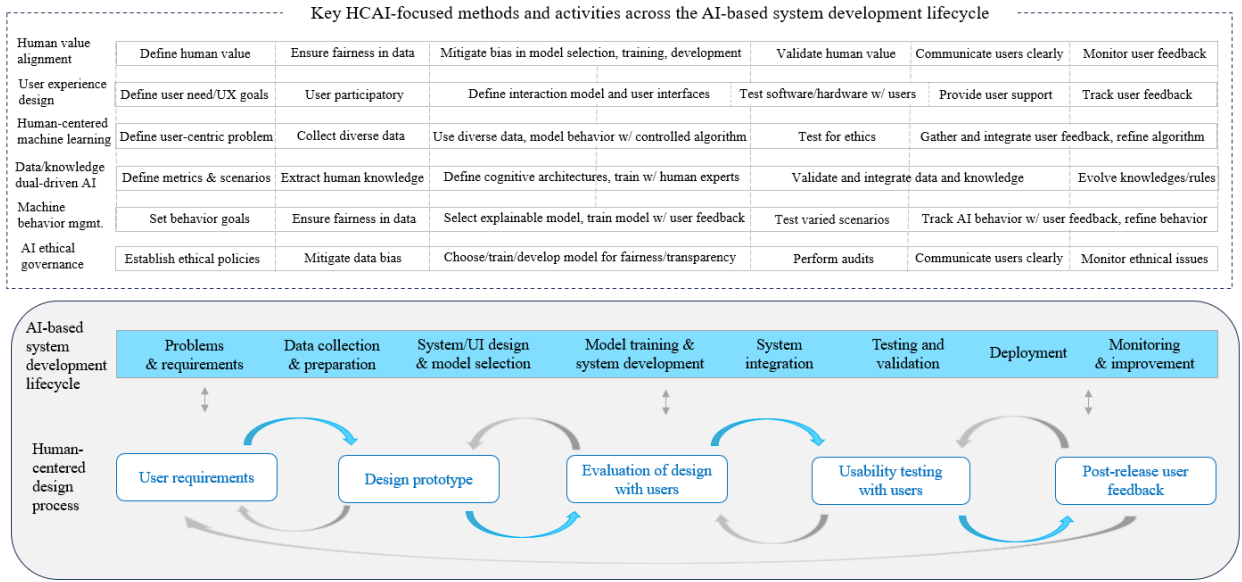

Figure 5  Illustration of the human-centered process for HC-HAII

As shown in Figure 5 (blue box), the full lifecycle of AI systems generally follows a process from "Problems & Requirements" to "Monitoring & Improvement". On this basis, the human-centered process for HC-HAII systematically integrates the AI-based system development lifecycle and the HCD process. The human-centered process for HC-HAII spans the entire AI lifecycle, encompassing the clarification of HCAI design goals and principles in the early stages, the design and development of effective human-AI interactions based on HCAI, and subsequent activities such as AI governance based on HCAI.

In the human-centered process, as illustrated in Figure 5, a series of methods proposed in this book are integrated into each stage, forming an "end-to-end" process based on the HCAI approach. This process represents a working mindset based on the HCAI approach, facilitating the delivery of human-centered AI systems.

## 4.3 Human-centered multi-disciplinary collaboration

A fundamental driving force for technological development stems from interdisciplinary collaboration. The AI development itself also benefits from interdisciplinary collaboration. As discussed earlier, AI technology introduces new characteristics distinct from traditional non-AI computing technology, bringing a series of new challenges to HAII. To find effective solutions to address these challenges, Table 6 compares the strengths and weaknesses of AI-related disciplines with human-centered disciplines, such as human factors and human-computer interaction in addressing these new challenges (Xu & Dainoff, 2022; Xu, Dainoff, Ge, & Gao, 2023). As shown in Table 6, the series of new challenges brought by AI technology poses challenges to HAII research and applications. Importantly, although AI disciplines and human-centered disciplines have their strengths and weaknesses, they complement each other, which explains the necessity of interdisciplinary collaboration. Multidisciplinary collaboration will enable the development of human-centered AI systems more effectively.

Table 6  Comparison Between AI-Related Disciplines and Human-Centered Disciplines
in Addressing Emerging Challenges

| Transformative Characteristics | New Challenges Introduced by AI Technologies | Pros and Cons of AI-Related Disciplines in Addressing the Challenges | Pros and Cons of Human-Centered Disciplines in Addressing the Challenges |
|---|---|---|---|
| From expected to unexpected machine behavior | • Unique autonomous capabilities (e.g., self-learning, adaptation, self-execution) may handle some scenarios that designers cannot anticipate<br>• AI systems exhibit unique behaviors and indeterministic outputs, causing potentially biased outcomes | Pros<br>• Handle some unexpected scenarios in emergencies<br>• Enable systems to be more productive than individual humans based on AI technologies (e.g., big data, machine learning)<br>Cons<br>• Lack of methods informed by behavioral science for managing machine behavior<br>• Do not consider evolving system behavior for testing/validation in the current software engineering approach | Pros<br>• Conduct iterative design prototyping and testing activities to help with data collection, training, and testing for optimizing algorithms<br>• Collect system and user feedback to support continued improvement of machine behavior<br>• Deploy user-participatory design to support human-centered and interactive machine learning<br>Cons<br>• Design relies on anticipated system behaviors and is hard to deal with abnormal/unexpected scenarios |
| From human-computer interaction to human-AI collaborative interaction | • AI agents with autonomous capabilities may collaboratively interact with humans | Pros<br>• Develop AI agents to interact with humans collaboratively<br>• Enhance overall system performance through human-AI collaborative interaction<br>Cons<br>• Lack of cognitive models and theories to implement human-AI collaborative interaction<br>• Primarily adopt a technology-centered approach (i.e., humans adapt to AI) | Pros<br>• Leverage existing human-human theories for modeling human-AI collaborative interaction<br>• Help measure human-AI collaborative interaction with current HCI evaluation methods<br>• Primarily adopt a human-centered approach<br>Cons<br>• Lack mature models for human-AI collaborative interaction (e.g., mutual trust and situation awareness, shared control)<br>• Focus on human interaction with non-AI systems in current approaches |
| From siloed human intelligence to hybrid intelligence | • Machines possess some cognitive capabilities like humans (e.g., learning, reasoning) | Pros<br>• Develop hybrid intelligence to produce more powerful intelligence by taking the complementary advantages of human and machine intelligence<br>Cons<br>• Hard to emulate advanced human cognitive capabilities, resulting in challenges in developing machine intelligence in isolation<br>• May not guarantee human-centered design in hybrid intelligence (i.e., humans may lose ultimate decision-making authority) | Pros<br>• Advocate human-controlled hybrid intelligence<br>• Potential to help model advanced human cognitive capabilities<br>Cons<br>• Lack methods for modeling advanced human cognition<br>• Lack of consideration of machine intelligence in system design |
| From usability to explainability (XAI) | • The AI black box effect may make the output obscure to users, affecting user trust and acceptance | Pros<br>• Develop transparent and explainable algorithms<br>• Develop explainable UI models with advanced HCI technology<br>Cons<br>• Generate more complex algorithms with an algorithm-driven approach, resulting in less explainable AI<br>• Lack of participation from other disciplines<br>• Lack of validated psychological models for explanation | Pros<br>• Help deliver XAI with advanced UI and visualization technologies<br>• Help accelerate the transfer of psychological theories<br>• Build human-centered XAI through user-exploratory and user-participation approaches<br>• Help validate XAI with HCI testing methods<br>Cons<br>• Focus on system usability in current HCI approaches |
| From automation to autonomy | • Different from automation (based on fixed rules or logics), autonomous systems may produce indeterministic results and can handle some unexpected situations<br>• The public's perception and overtrust of autonomous technology may cause safety issues<br>• Confusion between automation and autonomy may lead to inappropriate expectations and potential misuse of technology | Pros<br>• Develop autonomous systems to handle some unexpected scenarios in emergencies that could not be handled using automation<br>Cons<br>• Develop systems that may be without ultimate human control due to a lack of an HCAI approach | Cons<br>• Transfer the lessons learned from human factors/HCI research on automation to avoid issues (e.g., overtrust, human "out-of-the-loop")<br>• Develop innovative research/design paradigm (e.g., collaborative human-autonomy teaming)<br>Cons<br>• Lack of an effective design paradigm that can enable human operators to monitor and quickly take over control of autonomous systems in emergencies |
| From conventional interactions to intelligent interactions | • Different from the traditional stimulus-response paradigm, intelligent systems can initiate proactive interaction<br>• Paradigmatic changes emerge for interaction design (e.g., user intent detection, affective interaction, fuzzy reasoning interaction) | Pros<br>• Develop intelligent interaction technologies with new technologies (e.g., AI, big data, sensing technologies)<br>Cons<br>• Lack of effective design paradigms and metaphors for intelligent interactions<br>• Cause an overload of human cognitive resources in a pervasive computing environment (e.g., ambient intelligence) | Pros<br>• Drive effective design paradigms and metaphors for intelligent interactions<br>• Design/develop natural and usable intelligent UI<br>• Adapt AI technology to human capabilities by optimizing the match of human cognitive workload and interaction technology<br>Cons<br>• Focus on traditional interaction design in current HCI design standards<br>• Late participation of HCI professionals in the development of intelligent systems |
| | | | • Cannot effectively support the development of intelligent interactions with current HCI design and prototyping methods |
| From general user needs to specific ethical AI needs | • New user needs are emerging (e.g., privacy, fairness, ethics) | Pros<br>• Increased awareness of ethical AI issues in developing AI systems<br>• Many ethical AI guidelines have been published<br>Cons<br>• Lack effective methods to translate ethical AI principles into practice<br>• Ethical design may be ad hoc activities after development<br>• Lack of technical examples for ethical design<br>• AI engineers typically lack formal training in applying ethics to design and tend to view ethical design as another form of solving technical problems | Pros<br>• Leverage methods (e.g., iterative design and testing) to translate user needs into the process of data collection as well as training, optimizing, and testing algorithms and machine behaviors<br>• Leverage HCI interdisciplinary skills to address ethical AI issues by adopting social and behavioral science methods<br>Cons<br>• Lack of best-known methods for addressing ethical AI |

As an interdisciplinary field, HC-HAII research and applications require support from multiple disciplines. Professionals engaged in HC-HAII, including those from AI, computer science, data science, human-computer interaction, human factors/ergonomics, engineering design, industrial design, cognitive neuroscience, social sciences, and anyone involved in the development and implementation of human-centered AI systems, fall within this scope. The interdisciplinary field also requires the involvement of those who work on the formulation and governance of AI ethical norms. The authors of this book come from different disciplinary fields, which fully reflects the interdisciplinary characteristics of HC-HAII research and applications.

## 4.4 Human-centered multi-level design paradigms

A design paradigm provides a specific lens that shapes the perspective, scope, approaches, and methods for a field. In the PC era, various design paradigms were employed to address challenges. For instance, human factors science applies the paradigm of human cognitive information processing, studying how psychological activities, such as perception, workload, and situation awareness, interact with systems (Wickens et al., 2021). Similarly, fields such as HCI and user experience treat computing systems as tools to support users, focusing on user models, UI concepts, and usability testing to enhance user experience (Xu, 2003).

As an emerging field, HAII needs effective design paradigms to tackle AI-related challenges in the AI era. New characteristics of AI technology and the new type of human-machine relationship based on human-AI collaborative interaction inevitably bring new thinking about research and design for HAII. To address these challenges, Xu (2022a, 2022b) proposed three frameworks of human-AI joint cognitive systems, human-AI joint cognitive ecosystems, and intelligent sociotechnical systems frameworks. This chapter extends the three frameworks to align with the human-centered multi-level design paradigms for HC-HAII (Xu, Gao, Ge, 2024; Xu, Gao, 2025). This extension offers fresh design thinking to HC-HCAI practice.

### 4.4.1 Human-AI Joint Cognitive Systems

As discussed earlier, human-AI collaborative interaction emerges as a new form of human-machine relationship in the AI era. Xu (2022a) proposed a conceptual framework for human-AI joint cognitive systems (HAJCS) to represent the human-AI collaborative interaction (Hollnagel & Woods, 2005; Xu, 2022a) (see Figure 6 and Table 7).

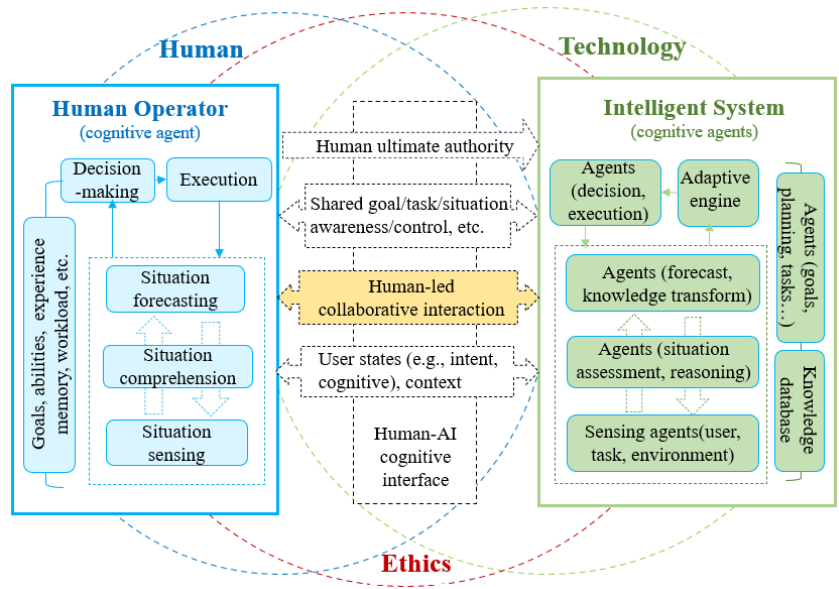

Figure 6 A framework for human-AI joint cognitive systems (HAJCS)

Table 7    Core Approaches and Focus Areas of Human-AI Joint Cognitive Systems

| Approach | Key Focus |
|---|---|
| 1. Human-centeredness | Emphasizes humans lead the collaborative interaction, with humans retaining final decision-making authority, especially in critical contexts. Integrates technology, human factors, and ethics in system design (illustrated by the three dotted-line circles). |
| 2. Machine Cognition Empowerment | Emulates AI agents with human-inspired cognition capabilities, enabling more capable AI technology while ensuring human oversight and controllability. |
| 3. Human-Led Collaborative Interaction with AI | Refines the human-AI relationship as a collaborative interaction between two cognitive agents, maintaining human leadership roles; It applies human team theories to develop collaborative interaction models and emphasizes the complementarity and mutual enhancement of human and AI intelligence within joint cognitive systems. |
| 4. Bi-directional Proactive Interaction | Advocates for mutual, two-way interaction where both AI and human agents recognize each other's states (e.g., physiological, cognitive, emotional), enhance mutual trust and situation awareness, and adapt accordingly through proactive and multimodal communication, while preserving human leadership and control. |

As shown in Figure 6 and Table 7, unlike traditional human-computer interaction systems, this framework views the AI system (with one or more AI agents) as a cognitive agent capable of performing some cognitive tasks as enabled by AI technology. Therefore, a human-AI system can be represented as a joint cognitive system in which two cognitive agents interact collaboratively. Depending on the level of advancement, an AI system can autonomously perceive, recognize, learn, and reason about the user's state, the environmental context, and other relevant factors, and subsequently carry out appropriate autonomous actions (Kaber, 2018; Xu, 2020).

This framework utilizes Endsley's situation awareness cognitive theory to represent the information processing mechanisms of human and machine cognitive agents (Endsley, 1995, 2015), specifically the cognitive processing mechanism of human operators in perceiving, understanding the current environmental state, and predicting future states. As shown in Figure 6, this model uses a similar cognitive mechanism to represent the information processing mechanism of the machine cognitive agent. The human-AI joint cognitive system framework presents a novel design paradigm for HAII, primarily characterized by the following new features.

This design paradigm can be applied to a wide range of application scenarios where both agents collaboratively interact while maintaining human controllability (refer to the corresponding chapters of this book), for example:

- *Chapter 8: Brain-computer interface and brain-computer fusion.* A human-AI joint cognitive system in the context of brain-computer interface/fusion, where humans and AI agents interact collaboratively through hybrid intelligence, integrating human biological intelligence and machine intelligence.
- *Chapter 22: Human-centered collaborative automated driving.* A human-AI joint cognitive system in the context of collaborative co-driving, where human drivers and onboard AI agents work together to accomplish driving tasks.
- *Chapter 26: Pilot-centered intelligent civil aircraft cockpit.* A human-AI joint cognitive system in the context of single pilot operation (SPO) of large civil aircraft, where a human pilot and airborne AI agent (co-pilot) interact collaboratively with dynamic functional and task allocations across flight scenarios.

### 4.4.2 Human-AI Joint Cognitive Ecosystems

The design paradigm based on the human-AI joint cognitive system framework mainly targets individual human-AI systems. More AI-based intelligent ecosystems, composed of multiple AI systems, are gradually forming, such as smart cities, intelligent transportation, and intelligent healthcare. In these intelligent ecosystems, HAII exists not only in the form of individual human-AI systems but also occurs across multiple human-AI systems. The safety and performance of an entire intelligent ecosystem depend on the coordination and collaborative interaction across multiple human-AI systems within the ecosystem.

For example, the entire human-car co-driving intelligent ecosystem encompasses collaborative interactions among humans, cars, intelligent traffic systems, and traffic command centers. The collaborative interactions across these individual human-AI joint cognitive systems will directly affect the driving safety of the entire intelligent transportation ecosystem. Therefore, HAII research and applications need to consider systematic solutions for the entire human-AI joint cognitive ecosystem from an ecosystem perspective.

Currently, research on multiple AI systems primarily focuses on engineering technology; there is a lack of research that considers the overall design of the system from the HAII perspective. Xu (2022a) initially proposed a conceptual framework for characterizing the intelligent ecosystem as a human-AI joint cognitive ecosystem (see Figure 7 and Table 8).

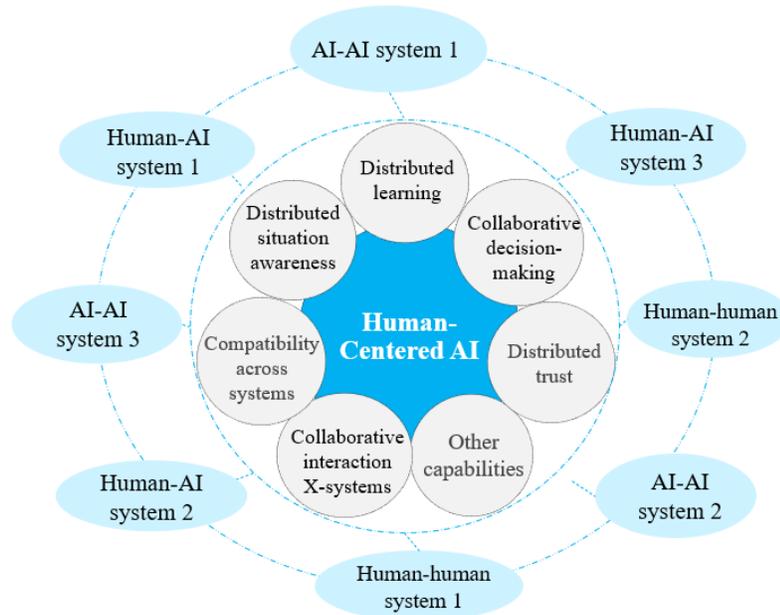

Figure 7 A conceptual framework of Human-AI joint cognitive ecosystem

Table 8  Core Approaches and Focus Areas of Human-AI Joint Cognitive Ecosystems

| Approach | Key Focus |
|---|---|
| 1. Systematic Design | Treats an intelligent ecosystem as a network of interconnected human-AI systems, requiring coordination beyond individual system optimization. Shifts focus from isolated systems to system-wide coordination of ecosystems. Emphasizes design coherence across all human-AI joint cognitive systems within intelligent ecosystems. |
| 2. Human-Centered Design | Prioritizes human roles, values, and authority within the ecosystem to ensure trust, accountability, and ethical alignment across human-AI systems. Ensures human oversight and control in decision-making. Addresses potential conflicts across systems with different societal or cultural origins. Reinforces trust and human primacy. |
| 3. Distributed Coordination and Collaborative Interaction | Emphasizes networked collaborative interactions among subsystems through shared situation awareness, decision-making, and emotional and social intelligence. Promotes seamless coordination in complex, multi-agent environments through trust, cognition, emotion, and communication across distributed human-AI systems. |
| 4. Continuous Learning and Evolution | Supports adaptive and sustainable growth of the ecosystem through co-learning, knowledge exchange, and self-organization. Leverages human and AI complementarity for co-evolution and dynamic adaptation to complex and changing eco-environments. |

As shown in Figure 7 and Table 8, an intelligent ecosystem can be represented as a *human-AI joint cognitive ecosystem*, composed of multiple human-AI subsystems (human-AI joint cognitive systems). For instance, the intelligent transportation ecosystem includes human–vehicle, human–human, vehicle–vehicle, and vehicle–environment interactions. The overall performance and safety of such an ecosystem rely on the integrated optimization of these collaborative cognitive subsystems. This ecosystem-based framework introduces a new design paradigm for HAII, characterized by distinct features and emerging research directions.

The framework of the human-AI joint cognitive ecosystem provides a new perspective for HAII research and applications. Currently, the development of intelligent ecosystems is just beginning, and there is a need to develop a series of theories and methods for human-AI joint cognitive ecosystems. This book presents several application scenarios of intelligent ecosystems, such as intelligent healthcare, intelligent manufacturing, and intelligent transportation.

- *Chapter 22: Human-centered collaborative automated driving.* A human-AI joint cognitive ecosystem in the context of autonomous driving, where the ecosystem brings together human drivers, in-vehicle AI systems, vehicles, and infrastructure to enable safe and effective co-driving.
- *Chapter 23: Patient-centered intelligent healthcare.* A human-AI joint cognitive ecosystem in the context of intelligent healthcare, where the ecosystem connects people (e.g., patients, physicians), AI tools, data, and intelligent healthcare systems to deliver better, more personalized care.
- *Chapter 27: Human-centered intelligent manufacturing.* A human-AI joint cognitive ecosystem in the context of intelligent manufacturing, where the ecosystem combines people (e.g., engineers, managers, planners), intelligent machines (e.g., sensors, robots, connected devices), and digital technologies to make production faster, safer, and more flexible.

### 4.4.3 Intelligent Sociotechnical Systems

AI systems and intelligent ecosystems are developed and used within broader sociotechnical environments. Sociotechnical systems (STS) theory emphasizes the joint optimization between technical, social, and organizational subsystems to achieve optimal system performance (Eason, 2011). This principle applies equally to modern AI systems.

Historically, human-machine system design focuses on physical user interfaces and environments, often overlooking the broader social and organizational context. However, the new challenges in the AI era, such as impacts on privacy, ethics, decision-making authority, work system (job roles, task flows, etc.), and organizational structures, require macro-level system consideration (Stahl, 2021).

Current STS theories do not fully address the complexities of AI systems. In response, Xu (2022b) proposed an Intelligent Sociotechnical Systems (iSTS) framework (see Figure 8 and Table 9). Compared with human-AI joint cognitive ecosystems, iSTS pays more attention to the impact of macro and non-technical factors on the design, development, deployment, and use of AI systems, including the redesign of work and life systems, organizational culture, organizational decision-making flow, etc. iSTS provides a new design paradigm for HAII from the broad sociotechnical perspective (Xu & Gao, 2025).

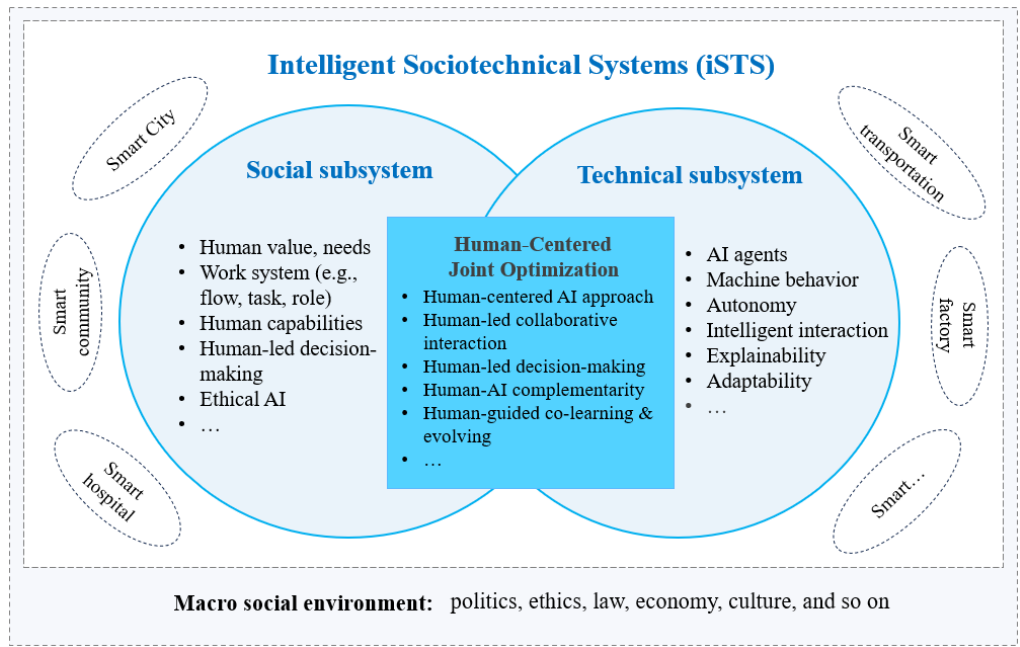

Figure 8 A Conceptual Framework of Intelligent Sociotechnical Systems (iSTS)

Table 9   Core Approaches and Focus Areas of Intelligent Sociotechnical Systems

| Approach | Key Focus |
|---|---|
| 1. Systematic Design Thinking | Integrates technical and non-technical factors (e.g., culture, social norms, organizational structures) into system design. Emphasizes multidimensional design within a sociotechnical context. |
| 2. Human-Centered Design | Places human needs, values, and roles at the core of AI system design. Promotes ethical alignment, user participation, and preservation of human authority. Aims to augment human capabilities, not replace them. |
| 3. Integrated social agents | Views AI as integrated social agents within organizational and social systems. Emphasizes shared goals, mutual trust, and coordinated human-AI actions. |
| 4. Organizational Adaptation and Redesign | Recognizes the organizational disruption caused by AI and the need for redesign. Advocates for structural changes to support human-AI interaction, balancing autonomy, ensuring fairness, and protecting employee well-being. |
| 5. Human-AI Co-Learning and Co-Adaptation | Supports mutual learning and behavioral evolution between humans and AI. Promotes co-evolution of humans and AI. Fosters governance mechanisms to sustain human value alignment and performance over time. |
| 6. Open Ecosystem Perspective | Moves beyond fixed boundaries in traditional sociotechnical systems. Addresses ethical, design, and human value alignment challenges in open, dynamic environments. |

The iSTS framework offers a new paradigm for HAII, emphasizing sociotechnical integration. It encourages interdisciplinary approaches that consider ethics, governance, future intelligent social forms, and human-centered applications such as intelligent healthcare, intelligent manufacturing, intelligent education, and AI-based assistive technology. Please refer to the following chapters.

- *Chapter 21: Human-centered approach for AI-based rehabilitation technology and accessibility design.* Rather than focusing solely on technical performance, iSTS emphasizes the co-adaptive interaction between humans and AI, promoting personalized, assistive solutions that respect individual needs, capabilities, and goals. It supports participatory design, ethical alignment, and contextual sensitivity, ensuring that rehabilitation technologies and accessibility tools enhance human dignity and quality of life.

- *Chapter 23: Patient-centered intelligent healthcare.* This approach ensures that AI-driven solutions are not only technically effective but also aligned with patients' values, needs, and preferences. By fostering collaboration among clinicians, patients, AI systems, and healthcare infrastructure, it enhances trust, improves care coordination, and empowers patients as active participants in their health journeys.
- *Chapter 27: Human-centered intelligent manufacturing.* It integrates AI with the human, organizational, and environmental aspects of production systems. Rather than viewing automation and AI as replacements for human labor, it promotes flexibility, resilience, and ethical alignment in manufacturing environments by enabling human-AI co-working, dynamic task allocation, and continuous learning, so that manufacturing systems are not only efficient and intelligent but also sustainable, inclusive, and centered on human values.

### 4.4.4 Implications of multiple design paradigms

Figure 9 illustrates the relationship among the three HC-HAII design paradigms, which represent an evolutionary expansion from a "1-D point" (i.e., focusing on individual human-AI systems) to a "2-D plane" that encompasses multiple human-AI systems, human-human interactions, and AI system networks, and ultimately to a "3-D space", reflecting a fully integrated intelligent sociotechnical environment. This is a comprehensive new design thinking approach that helps develop human-centered AI systems systematically and thoroughly. Figure 10 further compares the scope and characteristics of the three design paradigms.

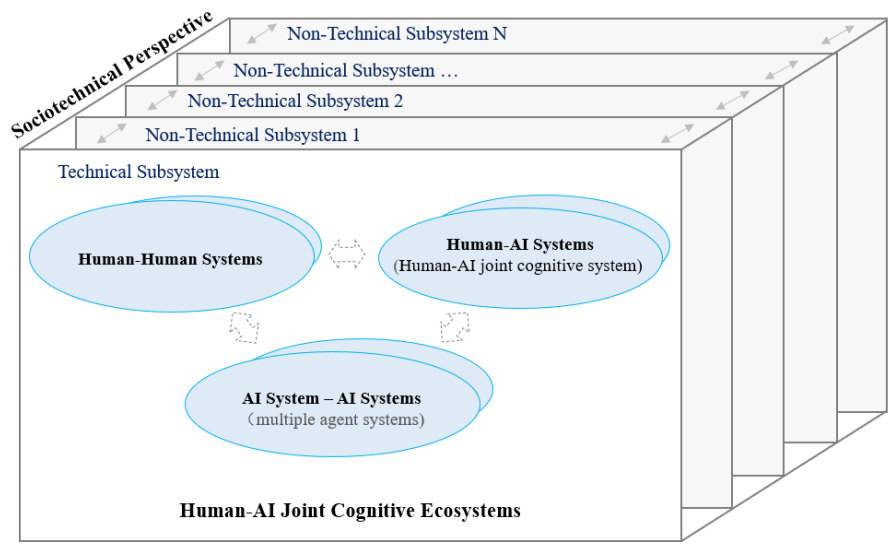

Figure 9 The relationship among the three human-centered HAII design paradigms

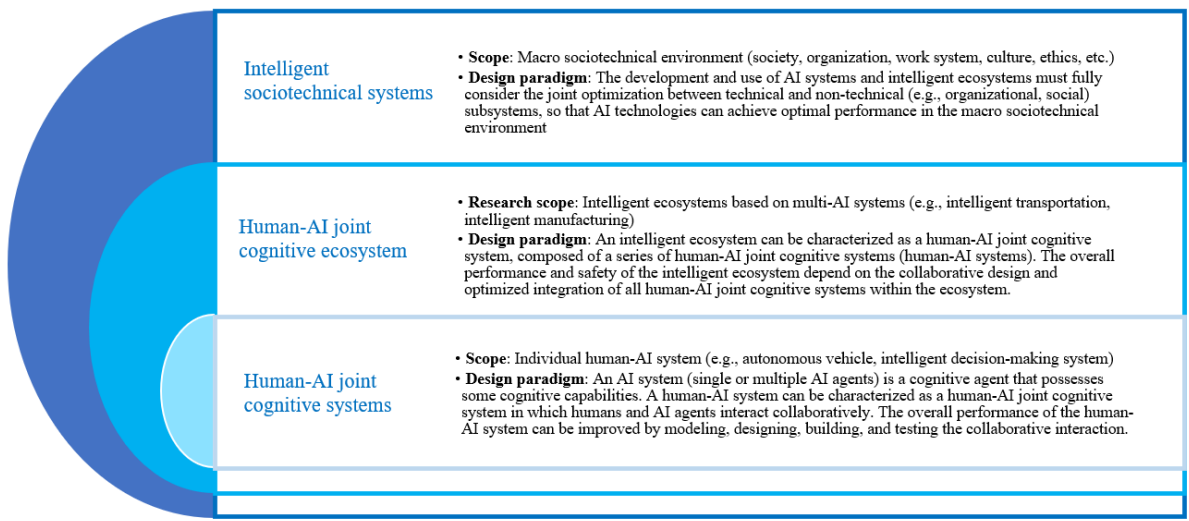

Figure 10 The scope and characteristics of the three human-centered HAII design paradigms

HAII research and applications require diverse design paradigms. The three design paradigms complement existing paradigms. Table 10 summarizes the significance of existing HAII design paradigms (e.g., computer science and AI disciplines, human factors science) and the three HC-HAII design paradigms. Among them, the column headings are for the diverse paradigms of HAII research and applications, and the row headings represent key HAII areas covered in this book. The cells summarize a series of unique HAII issues that need to be addressed by these diverse design paradigms.

Table 10   Implications of design paradigms for HAII research and applications

| Research area | Computational approach (AI/CS) | Human-centered design （human factors science） | Human-AI joint cognitive systems | Human-AI joint cognitive ecosystems | Intelligent sociotechnical systems |
|---|---|---|---|---|---|
| AI-based machine behavior | Algorithm optimization to minimize system bias | Apply user-participatory test methods in algorithm training and tuning to minimize bias | The impact of potential human-AI collaboration on machine behavior (e.g., co-learning, co-evolving) | Machine learning and behavior evolution across multiple AI systems within human-AI ecosystems | Effects of social factors on machine behavior and the ethics of machine behavior |
| Huma-led collaborative interaction with AI | Computational models of recognizing user states (e.g., intent, emotion) | Collaboration-based cognitive interface model, design, and user validation | Exploration of human-led collaborative interaction theory (e.g., mutual trust, shared control, and human leadership) | Collaborative interaction, adaptation, co-learning, and evolution mechanisms across multiple AI systems | Social impacts of human-AI collaboration and social responsibility on collaborative interaction |
| Human-machine hybrid intelligence | Advanced human cognition modeling, "data + knowledge" dual-driven AI | The "human-in/on-the-loop" interaction design: complementary approaches to leveraging human-AI advantages | Models and methods of human-AI hybrid intelligence based on human-AI collaborative interaction | Distributed collaborative interaction, cognition, and human-AI co-learning theories across multiple AI systems | Human-AI function allocation and human authority setting in the social environment |
| Ethical AI | Algorithm governance, algorithm training & optimization, reusable ethical AI codes | Human controllable AI design, meaningful human control, and behavioral methods for ethical AI design | Ethical issues of human-AI collaboration (e.g., norms, authority, culture) | Human authority setting and compatibility across multiple AI systems due to different cultures and norms | Ethical AI issues in the social environment, ethical AI norms, and governance |
| Human-AI interaction | Algorithms for intelligent interaction (e.g., emotion, intent) | New UI design paradigms for AI systems, HAII design standards | Cognitive UI and interaction design paradigms based on human-AI collaboration | Interaction compatibility theory, technology, and design across multiple AI systems | Impacts of social environments (cultural, ethical, etc.) on interaction design |
| Explainable AI (XAI) | Explainable AI models and algorithms | Human-centered explainable AI methods, psychological explanation theory transformation, explainable AI UI models, and visualization design | Team-Aware explainability, contextualized and role-specific explanations, and team performance of XAI | Explainable AI issues across AI systems within an ecosystem (e.g., compatibility and conflict management across AI systems) | Public AI trust and acceptance, and the relationship with AI interpretability; the relationship between explainable AI, culture, user knowledge, and ethics |
| Intelligent healthcare | Algorithms and technologies, intelligent medical system | Interaction design of intelligent healthcare systems, explainability of AI systems, holistic user experience (e.g., doctors, patients) | Collaborative decision-making of intelligent medical systems across roles (e.g., clinical doctors, pathologists, patients) | Coordination and collaboration within an intelligent medical ecosystem (e.g., manufacturers, insurance companies, hospitals, clinical doctors, patients) | Macro social and organizational factors affecting intelligent healthcare, patient trust, data privacy |

As seen in Table 10, HAII research and applications require diverse design paradigms. Existing design paradigms still play a significant role, and the three HC-HAII-driven design paradigms proposed in this chapter complement the existing paradigms. These new paradigms help broaden HAII scope, break out of the existing design thinking, carry out comprehensive HAII practice, and effectively solve the new challenges of AI technology to achieve HCAI solutions. Emerging AI technologies have expanded the existing human-computer interaction (HCI) design paradigms, and HAII research and applications continue to further evolve these design paradigms. The success of HAII research and applications requires diverse, interdisciplinary, and progressive design paradigms.

## 5 Structure of this book

This book takes the HCAI approach as its primary focus and comprehensively introduces the latest research and applications in HAII, organized into the following sections.

*Overview.* The "Overview" section introduces the conceptual foundations, scope, and structure of the book. It sets the stage by defining human-centered AI (HCAI) and situating human-AI interaction (HAII) from the HCAI perspective. Then it proposes a human-centered HAII (HC-HAII) framework, emphasizing the needs and presenting the methodology for HAII research and applications from the HCAI perspective. This section outlines the book's organization, introduces key emerging themes in HAII, and provides readers with a comprehensive understanding of the book's objectives. It also maps how the chapters align across theoretical foundations, technical innovations, design strategies, and practical applications, offering a roadmap for readers to navigate the evolving field of HAII from the HCAI perspective.

*Conceptual Foundations.* The "Conceptual Foundations" section introduces theoretical and methodological perspectives that underpin the HAII field from the HCAI perspective. It brings together diverse chapters that examine the cognitive, behavioral, and human factors dimensions essential for understanding how AI systems can effectively support, complement, and interact with humans. Topics such as computational modeling of human cognition, dual-driven AI frameworks, human state recognition, and meaningful human control offer foundational insights that inform the design and implementation of AI systems. These chapters support more advanced discussions in subsequent sections of the book, helping readers grasp the human-centered principles that shape this interdisciplinary domain.

*Human-AI Interaction Technology and Research.* This section presents five chapters that explore emerging technologies and research directions in HC-HAII, supporting richer and more immersive human-AI interaction. Topics include brain-computer interfaces and brain-computer fusion, which open new modalities for human intent detection and system response, and the human-centered human-AI collaboration (HCHAC) framework, which proposes a new structure for possible human-AI collaboration grounded in human-centered values. Chapters also address multimodal interaction via wearable devices, emphasizing real-time, context-aware feedback loops. Additional contributions investigate novel input modalities, interactive system behaviors, and cognitive augmentation technologies. The section bridges human-computer interaction (HCI) traditions with new research and applications at the intersection of AI, neuroscience, and human factors.

*Human-AI Interaction in Computing.* The "Human-AI Interaction in Computing" section comprises five chapters that explore how computational mechanisms and algorithmic structures can be designed to align with human-centered objectives. These include human-centered explainable AI, which enhances system transparency and user understanding; human-centered perceptual computing, which captures sensory and contextual cues for adaptive and human-centered decision-making; and human-centered algorithmic nudging, which aims to influence user behavior in an ethical and supportive manner. The section also addresses cognitive load balancing and interactive computation in uncertain environments, highlighting the role of computational design in enhancing the intuitiveness, safety, and social acceptability of AI systems.

*Human-AI Interaction Design.* The "Human-AI Interaction Design" section comprises four chapters that focus on establishing design frameworks, interaction design paradigms, and modeling system behavior for HCAI systems. It begins with an overview of human-AI interaction design standards and principles that ensure transparency, fairness, and user experience. Chapters delve into theoretical models and interaction paradigms that guide how AI systems respond to human inputs and contexts, as well as how designers can shape these responses. One chapter addresses the critical role of machine behavior design and management, including how AI systems exhibit, adapt, or constrain behaviors in collaborative settings. Together, these chapters offer conceptual and methodological tools for building HCAI systems.

*Human-AI Interaction in Practice.* The "Human-AI Interaction in Practice" section presents six chapters that demonstrate how HCAI principles are applied to HAII research and applications across real-world domains. Topics include collaborative automated driving, where the interaction between human drivers and AI systems is optimized for safety. Other chapters explore patient-centered healthcare that personalizes diagnostics and treatment planning with AI, as well as AI-assisted tools for autism diagnosis that support therapists, children, and families in clinical settings. Additional chapters examine the use of AI in education, public health, aviation, and manufacturing environments, emphasizing socio-technical integration of AI technology. This section grounds the theoretical and technical discussions from earlier sections in practical, often multidisciplinary case studies.

Additionally, each chapter summarizes the challenges currently faced and the key issues that need to be addressed in the future. Therefore, this book not only introduces the status of the HAII from the HCAI perspective but also looks forward to future directions. Through these chapters, the book offers readers a comprehensive overview of the latest research and application advances in HAII from the HCAI perspective.

## 6 Conclusion

The emergence of human-centered AI (HCAI) marks a paradigm shift in how people conceptualize and design AI systems, placing humans at the core rather than at the periphery of AI systems with which humans interact daily. In the context of human-AI interaction (HAII), the HCAI approach is not merely a design philosophical stance but a practical necessity to ensure AI technologies are useful, usable, safe, ethical, and empowering. This chapter presents a guiding framework in the form of human-centered human-AI interaction (HC-HAII), emphasizing a human-centered approach for HAII research and applications. It discusses the distinctive challenges and future directions facing the field. As AI systems become increasingly autonomous, socially interactive, and embedded in complex human environments, the need for human-centered thinking will only intensify.

The chapter presents the HC-HAII methodology, including human-centered methods, process, interdisciplinary collaboration, and multi-level design paradigms, and highlights key research challenges and future directions. The chapter also provides a structural overview of this handbook, which brings together contributions from an interdisciplinary community of researchers and practitioners to advance the theory, methodology, and application of HCAI in diverse domains of HAII. The purpose of this chapter is to provide a basic framework for this book. The contributions in this handbook aim to provide a comprehensive foundation for advancing research and applications in HAII from the HCAI perspective, with the goal of advancing HAII research and applications for human-centered systems.

As an emerging interdisciplinary field, the theory and methods of HAII are still in development, requiring reference to theories and methods from other disciplines, as well as further refinement of theories and methods based on HCAI. As a new field, it is necessary to conduct in-depth research on some key issues, thereby providing a basis for comprehensively carrying out HAII research and applications. As an applied field, HAII research must demonstrate its application value and further promote the implementation of the HCAI approach and HAII practices.

From the broad perspective of HC-HAII, despite rapid progress in the advances of AI technologies, realizing the full potential of HCAI in HAII remains a significant challenge. As AI systems become increasingly embedded in critical aspects of human life, it is imperative to address both technical hurdles and sociotechnical complexities that impact human-centered AI systems. Significant challenges and future work, for example, include the integration of interdisciplinary methods, ethical alignment, responsible innovation, and human-AI collaborative interaction while maintaining human leadership. This book brings together diverse perspectives to illuminate these challenges and inspire actionable strategies for advancing the practice of HCAI in the evolving landscape of HAII.

## References

AIAAIC (AI, Algorithmic, and Automation Incident and Controversy) (2025). *AI Incident Documentation Practices*. [Online]. *Available:* https://www.aiaaic.org/. . [Accessed: April 26, 2025].

Allenby, B. R. (2021). World Wide Weird: Rise of the Cognitive Ecosystem. *Issues in Science and Technology*, 37(3), 34-45.

Amershi, S., Cakmak, M., Knox, W. B., & Kulesza, T. (2014). Power to the people: The role of humans in interactive machine learning. *AI Magazine*, 35(4), 105−120.

Amershi, S., Weld, D., Vorvoreanu, M., Fourney, A., Nushi, B., Collisson, P., ... & Horvitz, E. (2019, May). Guidelines for HAII. *In Proceedings of the 2019 chi conference on human factors in computing systems* (pp. 1-13).

Auernhammer, J. (2020). Human-centered AI: The role of human-centered design research in the development of AI. *In S. Boess & M.* Cheung (Eds.), *Synergy – DRS International Conference 2020.*

Badham, R., Clegg, C., Wall, T.(2000). Socio-technical theory. In: Karwowski, W.(Ed.), *Handbook of Ergonomics.* John Wiley, New York, NY.

Bansal, G., Nushi, B., Kamar, E., Lasecki, W. S., Weld, D. S., & Horvitz, E. (2019). Beyond accuracy: The role of mental models in human-AI team performance. *Proceedings of the AAAI Conference on Human Computation and Crowdsourcing*, 7(1), 2−11.

Bingley, W. J., et al. (n.d.). Where is the human in human-centered AI? Insights from developer priorities and user experiences. *Computers in Human Behavior.* vol. 141, p. 107617, 2023.

Brill, J.C., Cummings, M. L., Evans, A. W. *III.*, Hancock, P. A., Lyons, J. B. & Oden, K. (2018). Navigating the advent of human-machine teaming. *Proceedings of the Human Factors and Ergonomics Society 2018 Annual Meeting, 455-459.*

British Design Council. (2005). *The double diamond: A universally accepted depiction of the design process*. Design Council. Retrieved July 27, 2025, from https://www.designcouncil.org.uk/our-resources/the-double-diamond/

Budiu, R., & Laubheimer, P. (2018). *Intelligent assistants have poor usability: A user study of Alexa, Google Assistant, and Siri*. https://www.nngroup.com

Capel, T., & Brereton, M. (2023, April). What is human-centered about human-centered AI? A map of the research landscape. In *Proceedings of the 2023 CHI conference on human factors in computing systems* (pp. 1-23).

Cummings, M. L., & Clare, A. S. (2015). Holistic modelling for human-autonomous system interaction. *Theoretical Issues in Ergonomics Science,* 16(3), 214−231. https://doi.org/10.1080/1463922X.2014.1003990

den Broek, H.V., Schraagen, J. M., te Brake, G. & van Diggelen, J. (2017). Approaching full autonomy in the maritime domain: Paradigm choices and human factors challenges. *In Proceedings of the MTEC*, Singapore, 26-28 April 2017.

de Sio, S. F., & den Hoven, V. J. (2018). Meaningful human control over autonomous systems: A philosophical account. *Frontiers in Robotics and AI, 5*, 15. doi: 10.3389/frobt.2018.00015

Eason, K.(2011). Sociotechnical Systems Theory in the 21st Century: Another Half-filled Glass? Sense in Social Science: *A Collection of Essays in Honor of Dr. Lisl Klein.* Environmental Protection Agency.

Ehsan, U., Wintersberger, P., Liao, Q. V., Mara, M., Streit, M., Wachter, S., ... Riedl, M. O. (2021). Operationalizing human-centered perspectives in explainable AI. In Extended Abstracts of the 2021 *CHI Conference on Human Factors in Computing Systems* (pp. 1−6). Association for Computing Machinery.

Endsley, M. R. (2023). Ironies of artificial intelligence. *Ergonomics*, 1−13.

Endsley, M. R.(2015), Situation awareness misconceptions and misunderstandings. *Journal of Cognitive Engineering and Decision Making,* 9(1): 4-32.

Endsley, M.R.(1995).Toward a theory of situation awareness in dynamic systems. *Human Factors,* 37,32-64.

Floridi, L., et al. (2018). AI4People—An ethical framework for a good AI society: Opportunities, risks, principles, and recommendations. *Minds and Machines*, 28(4), 689−707. https://doi.org/10.1007/s11023-018-9482-5

Girardin, F., & Lathia, N. (2017). When user experience designers partner with data scientists. *In The AAAI Spring Symposium Series Technical Report: Designing the User Experience of Machine Learning Systems*. *The AAAI Press*. https://www.aaai.org/ocs/index.php/SSS/SSS17/paper/view/15364environment

Google PAIR. (2019). *People + AI Guidebook: Designing human-centered AI products.* Retrieved Nov. 23, 2023 from .

Hartikainen, M., et al. (2022). Human-centered AI design in reality: A study of developer companies' practices. *In Nordic Human-Computer Interaction Conference.*

Hollnagel, E., & Woods, D. D.(2005). *Joint cognitive systems: Foundations of cognitive systems engineering*. London: CRC Press. 2022, pp. 1−11.

Holmquist, L. E. (2017). Intelligence on tap: Artificial intelligence as a new design material. *Interactions*, 24(4), 28−33.

Hu, Y. D., Sun, X. H., Zhang, H. X., Zhang, S. C., & Yi, S. Q. (2020). Interaction design in human-in-the-loop hybrid intelligence. *Packaging Engineering*, 41(18), 38−47.

IEEE (The Institute of Electrical and Electronics Engineers). (2019). *Ethically aligned design: A vision for prioritizing human well-being with autonomous and intelligent systems*. The Institute of Electrical and Electronics Engineers (IEEE), Incorporated             .

ISO (International Organization for Standardization) (2019). *Ergonomics of human-system interaction Part 220: Processes for enabling, executing, and assessing human-centered design within organizations.* ISO 9241-220: 2019.

Kaber, D. B. (2018). A conceptual framework of autonomous and automated agents. *Theoretical Issues in Ergonomics Science*, 19(4), 406-430, DOI: 10.1080/1463922X.2017.1363314.

Kaluarachchi, T., Reis, A., & Nanayakkara, S. (2021). A review of recent deep learning approaches in human-centered machine learning. *Sensors*, 21(7), 2514.

Mazarakis, A., et al. (2023). What is critical for human-centered AI at work? Towards an interdisciplinary theory. *Frontiers in AI, 6, 1257057*.

McGregor, S. (2023). *AI Incident Database*. 2023. [Online]. Available: https://incidentdatabase.ai/. .[Accessed: Jul. 27, 2025].

Mueller, S. T., Hoffman, R. R., Clancey, W., Emrey, A., & Klein, G. (2019). Explanation in human-AI systems: A literature meta-review, synopsis of key ideas and publications, and bibliography for explainable AI. arXiv preprint arXiv:1902.01876.

NAS (National Academies of Sciences, Engineering, and Medicine). (2021). *Human-AI teaming: State-of-the-art and research needs*.

NITRD (2023). *The National Artificial Intelligence Research and Development Strategic Plan 2023 Update.* https://www.nitrd.gov/national-artificial-intelligence-research-and-development-strategic-plan-2023-update/

Rahwan, I., Cebrian, M., Obradovich, N., Bongard, J., Bonnefon, J.-F., Breazeal, C., ... & Wellman, M. (2019). *Machine behaviour*.Nature, 568(7753), 477-486.

Richards, D., Vythilingam, R., & Formosa, P. (2023). A principlist-based study of the ethical design and acceptability of artificial social agents. *International Journal of Human–Computer Studies*, 172, 102980.

Rong, Y., et al. (2024). Towards human-centered explainable AI: A survey of user studies for model explanations. *IEEE Transactions on Pattern Analysis and Machine Intelligence*, 46(4), 1–20.

Salehi, P., Chiou, E. K., & Wilkins, A. (2018, September). Human-agent interactions: Does accountability matter in interactive control automation?. *In Proceedings of the Human Factors and Ergonomics Society Annual Meeting* (Vol. 62, No. 1, pp. 1643-1647). Sage CA: Los Angeles, CA: SAGE Publications.

Santoni de Sio, F., & van den Hoven, J. (2018). Meaningful human control over autonomous systems: A philosophical account. *Front Robot AI*, 5(2), 15.

Schmager, S., Pappas, I. O., & Vassilakopoulou, P. (2025). Understanding Human-Centred AI: a review of its defining elements and a research agenda. *Behaviour & Information Technology*, 1-40.

Shively, R. J., Lachter, J, Brandt, S.L., Matessa, M., Battiste, V.& Johnson, W.W. (2018). Why human-autonomy teaming? *International Conference on Applied Human Factors and Ergonomics*, May 2018, DOI: 10.1007/978-3-319-60642-2_1.

Shneiderman, B. (2020). Human-centered artificial intelligence: Reliable, safe & trustworthy. *International Journal of Human–Computer Interaction*, *36*(6), 495-504.

Shneiderman, B. (2021) *19th Note: Human-Centered AI Google Group.* In Human-Centered AI (Sept.12, 2021). Available: https://groups.google.com/g/human-centered-ai/c/syqiC1juHO.c

Shneiderman, B. (2022). *Human-Centered AI*. Oxford University Press.

Stahl, B. C.(2021). *Artificial Intelligence for a Better Future: An Ecosystem Perspective on the Ethics of AI and Emerging Digital Technologies*(p. 124). Springer Nature.

Taddeo, M., Jones, P. et al. (2023). Sociotechnical ecosystem considerations: An emergent research agenda for AI in cybersecurity. *IEEE Transactions on Technology and Society*, vol. 4, no. 2, pp. 112–118, 2023.

Umbrello, S. & Natale, S. (2024). Reframing deception for human-centered AI. *International Journal of Social Robotics*, pp. 1–19, 2024.

van Allen, P. (2018). Prototyping ways of prototyping AI. *Interactions*, vol. 25, pp. 46–51, 2018.

Wickens, C. D., Helton, W. S., Hollands, J. G., & Banbury, S.(2021). *Engineering psychology and human performance*. Routledge.

Wickramasinghe, C.S. et al. (2020). Trustworthy AI development guidelines for human-system interaction, in *2020 13th International Conference on Human System Interaction (HSI)*, pp. 130–136.

Wortman, J., Wallach, H. (2021). A human-centered agenda for intelligible machine learning. In *Machines We Trust: Perspectives on Dependable AI*. Cambridge, MA, USA: MIT Press, 2021.

Xu, W. (2003). User-centered design: Challenges and opportunities for human factors. *Journal of Ergonomics*, *9*(4), 8-11.

Xu, W. (2019). User-centered design (III): Methods for user experience and innovative design in the intelligent era. *Chinese Journal of Applied Psychology*, 25(1), 3−17.

Xu, W. (2020). User-centered design (V): From automation to the autonomy and autonomous vehicles in the intelligence era. *Chinese Journal of Applied Psychology*, *26*(2), 108-129.

Xu, W. (2021). From automation to autonomy and autonomous vehicles: Challenges and opportunities for human-computer interaction. *Interactions*, *28*(1), 48-53.

Xu, W. (2022a). User-centered design (VI): Human factors engineering approaches for intelligent human-computer interaction. *Chinese Journal of Applied Psychology*, *28*(3), 191-209.

Xu, W. (2022b). User-Centered Design (VIII): A New Framework of Intelligent Sociotechnical Systems and Prospects for Future Human Factors Research, Chinese Journal of Applied Psychology, 28(5), 387-401.

Xu, W., Ge, L., & Gao, Z. (2021). Human-AI interaction: An emerging interdisciplinary domain for enabling human-centered AI. *CAAI Transactions on Intelligent Systems, 16*(4), 605-621.

Xu, W., Dainoff, M. J., Ge, L., & Gao, Z. (2022). Transitioning to human interaction with AI systems: New challenges and opportunities for HCI professionals to enable human-centered AI. *International Journal of Human-Computer Interaction*, 39(3), 494-518.

Xu, W., & Dainoff, M. (2023). Enabling human-centered AI: A new junction and shared journey between AI and Xu, W., Gao, Z. (2024). Applying human-centered AI in developing effective human-AI teaming: A perspective of human-AI joint cognitive systems. *Interactions*, 31 (1), 32-37.

Xu, W., Gao, Z., Dainoff, M. (2025).  An HCAI methodological framework (HCAI-MF): Putting it into action to enable human-centered AI. https://arxiv.org/abs/2311.16027

Yang, Q., Scuito, A., Zimmerman, J., Forlizzi, J., & Steinfeld, A. (2018). Investigating how experienced UX designers effectively work with machine learning. *In Proceedings of the 2018 Designing Interactive Systems Conference (DIS '18)* (pp. 585–596). *New York, NY: ACM.* https://doi.org/10.1145/3196709.3196730

Yang Q. *et al. (2020).* Re-examining whether, why, and how human-AI interaction is uniquely difficult to design. *CHI Conference on Human Factors in Computing Systems (CHI' 20)*, Honolulu, HI, USA, ACM, 2020, pp. 1–12.

Yampolskiy, R. V. (2019). Predicting future AI failures from historic examples. *Foresight*, vol. 21, pp. 138–152.

Yi, X., Yu, C., & Shi, Y. C. (2018). Bayesian method for intent prediction in pervasive computing environments. *Science China Information Sciences*, 48(4), 419−432.

Zanzotto, F. M. (2019). Human-in-the-loop artificial intelligence. *Journal of Artificial Intelligence Research*, 64(2), 243−252.

Zheng, N. N., Liu, Z. Y., Ren, P. J., Ma, Y. Q., Chen, S. T., Yu, S. Y., ... *Wang, F.* Y. (2017). Hybrid-augmented intelligence: Collaboration and cognition. *Frontiers of Information Technology & Electronic Engineering*, 18(2), 153−179.

Zeller, F. & Dwyer, L. (2022). Systems of collaboration: Challenges and solutions for interdisciplinary research in AI and social robotics. *Discover Artificial Intelligence*, vol. 2, no. 12.